\documentclass[onecolumn]{emulateapj}

\usepackage{color}
\usepackage{amsmath}
\usepackage{amssymb}

\renewcommand{\vec}[1]{\mbox{\boldmath$#1$}}
\renewcommand{\tensor}[1]{\mbox{\boldmath{\ensuremath{\mathsf{#1}}}}}

\shorttitle{Castro Radiation}
\shortauthors{Zhang et al.}

\begin{document}

\title{CASTRO: A New Compressible Astrophysical Solver. III. Multigroup
  Radiation Hydrodynamics}

\author{W.~Zhang\altaffilmark{1},
L.~Howell\altaffilmark{2},
A.~Almgren\altaffilmark{1},
A.~Burrows\altaffilmark{3},
J.~Dolence\altaffilmark{3},
J.~Bell\altaffilmark{1}}

\altaffiltext{1}{Center for Computational Sciences and Engineering,
                 Lawrence Berkeley National Laboratory,
                 Berkeley, CA 94720}

\altaffiltext{2}{Center for Applied Scientific Computing,
                 Lawrence Livermore National Laboratory,
                 Livermore, CA 94550}

\altaffiltext{3}{Dept. of Astrophysical Sciences,
                 Princeton University,
                 Princeton, NJ 08544}

\begin{abstract}
  We present a formulation for multigroup radiation hydrodynamics that
  is correct to order $O(v/c)$ using the comoving-frame approach and
  the flux-limited diffusion approximation.  We describe a numerical
  algorithm for solving the system, implemented in the compressible
  astrophysics code, CASTRO.  CASTRO uses an Eulerian grid with
  block-structured adaptive mesh refinement based on a nested
  hierarchy of logically-rectangular variable-sized grids with
  simultaneous refinement in both space and time.  In our multigroup
  radiation solver, the system is split into three parts, one part
  that couples the radiation and fluid in a hyperbolic subsystem,
  another part that advects the radiation in frequency space, and a
  parabolic part that evolves radiation diffusion and source-sink
  terms. The hyperbolic subsystem and the frequency space advection
  are solved explicitly with high-order Godunov schemes, whereas the
  parabolic part is solved implicitly with a first-order backward
  Euler method.  Our multigroup radiation solver works for both
  neutrino and photon radiation.

\end{abstract}

\section{Introduction}
\label{sec:intro}

Numerical simulations are a useful tool for many radiation
hydrodynamic problems in astrophysics.  However, numerical modeling of
radiation hydrodynamic phenomena is very challenging for a number of
reasons.  To achieve a highly accurate description of the system
usually requires a multidimensional treatment with high spatial
resolution.  It is also desirable to handle the radiation frequency
variable properly, because radiation transport is a
frequency-dependent process.  Moreover, the numerical algorithms for
radiation hydrodynamics must be stable and efficient.

Core-collapse supernovae are such complex phenomena \citep{Bethe90,
  KotakeST06, JankaLM07, Janka12} that neutrino radiation hydrodynamics
simulations have been indispensable for helping understand the
explosion mechanism.  Various algorithms have been developed to solve
neutrino radiation hydrodynamics equations.  Because of the
complexities involved in the problem and the limited computer
resources available, various tradeoffs have to be made.  Many of these
algorithms are 1D \citep[e.g.,][]{BowersWilson82, Bruenn85,
  MezzacappaBruenn93, BurrowsYP00, LiebendorferMM04, OConnorOtt10}, or
have used the so-called ``ray-by-ray'' approach, which does not truly
perform multi-dimensional transport \citep[e.g.][]{BurrowsHF95,
  RamppJanka02, BurasRJK06, TakiwakiKS12}.  The two-dimensional code
of \citet{Millerwm93} is a gray flux-limited diffusion (FLD) code that
does not include order $O(v/c)$ terms, where $v$ is the characteristic
velocity of the system.  The 2D multigroup multiangle code of
\citet{LivneBW04} does not include all order $O(v/c)$ terms, and in
particular it lacks direct coupling among radiation groups.  The
two-dimensional algorithm of \citet{HubenyBurrows07} is based on a
mixed-frame formulation of a multigroup two-moment system that is
correct to order $O(v/c)$, but it remains to be implemented in a
working code.  \citet{SwestyMyra09} have developed a fully coupled 2D
multigroup FLD code based on a comoving frame formulation that is
correct to order $O(v/c)$.  Besides the grid-based methods, 
smoothed particle hydrodynamics (SPH) methods have also been applied to
multi-dimensional neutrino radiation hydrodynamics simulations of
core-collapse supernovae \citep[e.g.,][]{HerantBH94, FryerWarren02,
  FryerRW06}, but the SPH scheme has not yet been extended to
multigroup.  Recently \citet{AbdikamalovBO12} developed a Monte
Carlo method for neutrino transport, but hydrodynamics and radiation
transport are not yet coupled in the scheme.

In this paper, we present a numerical algorithm for solving
multi-dimensional multigroup photon and neutrino radiation hydrodynamics
equations in the comoving frames that are correct to order $O(v/c)$.
Radiation quantities that are of interest are the radiation energy
density, radiation flux, and radiation pressure tensor.  It is more
accurate to evolve both the radiation energy density and the radiation
flux especially when the radiation is highly anisotropic and optically
thin.  However, the two-moment approach is very expensive in terms of
computer time and memory for multi-dimensional multigroup radiation
hydrodynamics simulations.  Thus, we have adopted the flux-limited
diffusion approach \citep{AlmeWilson73} for computational efficiency.
In the FLD approach, only the radiation energy density needs to be
evolved over time, and the radiation flux is derived through a
diffusion approximation. Furthermore, we use an analytic closure for
the radiation pressure tensor.  We have adopted a multigroup method,
in which the radiation frequency is discretized into multiple groups.
The equations we solve are similar to those in \citet{SwestyMyra09}.
However, we have neglected inelastic scattering of neutrinos on
matter, which is currently treated as elastic scattering.  Since the
total cross section for the latter at the neutrino energies near and
outside the neutrinospheres are $\sim 20$--100 times smaller than the
cross sections for scattering off nucleons and nuclei, not including
inelastic effects at this stage has a minor effect on the qualitative
behavior of collapse, bounce, and shock dynamics.  Nevertheless, the
approach of \citet{ThompsonBP03} to handle inelasticity and energy
redistribution can be incorporated into our scheme.  Another
difference between our scheme and the scheme of \citet{SwestyMyra09}
is that we have developed a Godunov scheme in which radiation is fully
coupled into a characteristic-based Riemann solver for the hyperbolic
part of the system.  

Many multigroup neutrino transport codes use a fixed Eulerian grid.
However, it is difficult to achieve sufficient resolution in
multi-dimensional simulations with this simple approach.  To
efficiently achieve high resolution, we use an Eulerian grid with
block-structured adaptive mesh refinement (AMR).  The AMR algorithm we
use has a nested hierarchy of logically-rectangular variable-sized
grids and refines simultaneously in both space and time.  AMR
techniques have been successfully used in multi-dimensional multigroup
photon radiation transport \citep{RAGE,crash}. However, to the best of
our knowledge, this is the first time that AMR has been employed in a
multi-dimensional multigroup neutrino solver.  Besides AMR, the
Arbitrary Lagrange Eulerian (ALE) method is an alternative approach
for mesh refinement.  We note that ALE has been successfully employed
in multi-dimensional multigroup neutrino transport
\citep{LivneBW04,BurrowsLD07,OttBDL08} and photon transport
\citep{HYDRA}.

We have implemented our algorithm in the compressible astrophysics
code, CASTRO.  This is the third paper in a series on the CASTRO code
and its numerical algorithms.  In our previous papers, we describe our
treatment of hydrodynamics, including gravity and nuclear reactions
\citep[][henceforth Paper I]{CASTRO}, and our algorithm for
flux-limited gray radiation hydrodynamics based on a mixed-frame
formulation \citep[][henceforth Paper II]{CASTRO2}.  Here, we describe
our algorithm for flux-limited multigroup radiation hydrodynamics
based on a comoving-frame formulation.  

In \S~\ref{sec:RHD}, we present the multigroup radiation hydrodynamics
equations that CASTRO solves.  We show that the system can be divided
into a hyperbolic subsystem (\S~\ref{sec:hyper}), a subsystem of
advection in frequency space (\S~\ref{sec:fspace}), and a parabolic
subsystem for radiation diffusion (\S~\ref{sec:para}).  Analytic
results for the mathematical characteristics of the hyperbolic subsystem
are also presented in \S~\ref{sec:hyper}.  The hyperbolic and the
frequency space advection steps are solved by an explicit solver,
whereas the parabolic subsystems is solved by an iterative implicit
solver.  We describe the explicit solvers in \S~\ref{sec:explicit},
followed by a description of the implicit solver in
\S~\ref{sec:implicit}.  We also discuss acceleration schemes that can
greatly improve the convergence rate of the iterative implicit solver
(\S~\ref{sec:accel}).  In \S~\ref{sec:tests}, we present results from
a series of test problems.  A scaling test is shown in
\S~\ref{sec:performance} to demonstrate the scaling behavior of
the multigroup group radiation solver.  Finally, the results of the
paper are summarized in \S~\ref{sec:sum}.

\section{Multigroup Radiation Hydrodynamics}
\label{sec:RHD}

In this section, we present the multigroup radiation hydrodynamics
equations that CASTRO solves.  Our formulation is based on a
comoving-frame approach, unlike our gray radiation hydrodynamics
solver presented in Paper II.  In the comoving-frame approach, the
radiation quantities and the opacities are measured in the comoving
frames.  The set of equations are correct to order $O(v/c)$.  We then
derive the multigroup radiation hydrodynamics equations using
frequency space discretization.  We also analyze the mathematical
characteristics of the system.  Here we present the results in terms
of neutrino radiation and indicate modifications needed for photon
radiation.

\subsection{Frequency-Dependent Radiation Hydrodynamics Equations}
\label{sec:fdrhd}

The system of neutrino radiation hydrodynamics that is correct to
order $O(v/c)$ can be described by 
\begin{align}
\frac{\partial \rho}{\partial t} + \nabla \cdot (\rho \vec{u}) = { } & 0, \label{eq:drhodt}\\
\frac{\partial (\rho \vec{u})}{\partial t} + \nabla \cdot (\rho \vec{u}
  \vec{u}) + \nabla p = { } & \int_0^{\infty} \frac{\chi}{c}
  \vec{F}_\nu \mathrm{d}\nu + \vec{F}_G, \label{eq:drhoudt} \\ 
\frac{\partial (\rho E)}{\partial t} + \nabla \cdot (\rho E \vec{u} + p
  \vec{u}) 
  = { } & \vec{u} \cdot \int_0^{\infty} \frac{\chi}{c} \vec{F}_\nu \mathrm{d}\nu +
  \int_0^{\infty} (c\kappa E_{\nu} - 4\pi\eta) \mathrm{d}\nu + \vec{u}
  \cdot \vec{F}_G, \label{eq:drhoEdt} \\
\frac{\partial (\rho Y_e)}{\partial t} + \nabla \cdot (\rho Y_e \vec{u}) 
 = { } &
\int_0^{\infty} \xi (c\kappa E_{\nu} - 4\pi\eta) \mathrm{d}\nu, \label{eq:drhoYedt} \\
\frac{\partial E_{\nu}}{\partial t} 
+ \nabla \cdot (E_{\nu} \vec{u})
= { } & -\nabla \cdot \vec{F}_\nu - (c\kappa E_{\nu} - 4\pi\eta) 
 + \frac{\partial}{\partial
  \ln{\nu}} (\tensor{P}_{\nu} : \nabla \vec{u}) , \label{eq:dEnudt}
\end{align}
\citep{Buchler83,MihalasMihalas99,Castor04,SwestyMyra09}. Here $\rho$,
$\vec{u}$, $p$, and $E$ are the mass density, velocity, pressure, and
total matter energy per unit mass (internal energy, $e$, plus kinetic
energy, $\vec{u}^2/2$), respectively. $Y_e$ is the electron fraction
defined to be the net electron number (electrons minus positrons) per
baryon. $\vec{F}_G$ is the external force on the fluid (e.g.,
gravitational force). $E_\nu$, $\vec{F}_\nu$, and $\tensor{P}_\nu$ are
the monochromatic radiation energy density, radiation flux, and
radiation pressure tensor at frequency $\nu$, measured in the comoving
fluid frame, respectively.  The speed of light is denoted by $c$. The
absorption coefficient is denoted by $\kappa$, and the total
interaction coefficient including conservative scattering is given by
$\chi$.  The emission coefficient is denoted by $\eta$.  For photon
radiation $\xi$ is zero, whereas for neutrinos $\xi$ is given by
\begin{equation}
  \xi = s \frac{m_{\mathrm{B}}}{h \nu},
\end{equation}
where $m_{\mathrm{B}}$ is the baryon mass, $h$ is the Planck constant,
and $s = 1$ for $\nu_e$ neutrinos, $s = -1$ for $\bar{\nu}_e$
neutrinos, and $s = 0$ for other flavor neutrinos.  Also note that for
neutrino radiation with multiple species, the integrals over frequency
imply summation over the species.  The colon product $:$ in
$\tensor{P}_{\nu} : \nabla \vec{u}$ indicates contraction over two
indices as follows,
\begin{equation}
  \tensor{P}_{\nu} : \nabla \vec{u} = 
  \left(\tensor{P}_{\nu}\right)_{ij}  \left(\nabla \vec{u}\right)_{ij}.
\end{equation}

The system of the radiation hydrodynamics (Equations
\ref{eq:drhodt}--\ref{eq:dEnudt}) is closed by an equation of state
(EOS) for matter, the diffusion approximation for radiation flux, and
a closure relation for radiation that computes the radiation pressure
tensor from radiation energy density and radiation flux.  In the FLD
approximation \citep{AlmeWilson73}, the monochromatic radiation flux
is written in the form of Fick's law of diffusion,
\begin{equation}
    \vec{F}_{\nu} = - \frac{c \lambda}{\chi} \nabla
    E_{\nu}, \label{eq:Fick} 
\end{equation}
where $\lambda$ is the flux limiter at frequency $\nu$.  We have
implemented a variety of choices for the flux limiter
\citep[e.g.,][]{Minerbo78, BruennBY78, LevermorePomraning81}.  For
example, one form of the flux limiter in \cite{LevermorePomraning81}
is
\begin{equation}
  \lambda = \frac{2+R}{6+3R+R^2}, \label{eq:lambda} 
\end{equation}
where
\begin{equation}
  R = \frac{|\nabla E_{\nu}|}{\chi E_{\nu}}. \label{eq:lam-R}
\end{equation}
In the optically thick limit, $R \rightarrow 0$, $\lambda \rightarrow
1/3$, and the flux approaches the classical Eddington approximation
$\vec{F}_\nu \approx - (c/3\chi) \nabla E_\nu$.  In the optically thin
limit, $R \rightarrow \infty$, $\lambda \approx 1/R$, and $\vec{F}_\nu
\approx c E_\nu$ as expected for streaming radiation.  For the
radiation pressure tensor, we use an approximate closure
\citep{Minerbo78, Levermore84}
\begin{equation}
    \tensor{P}_{\nu} = \frac{1}{2}(1-f) E_{\nu} \tensor{I} +
  \frac{1}{2}(3 f-1) E_{\nu} \hat{\vec{n}}\hat{\vec{n}},
\end{equation}
where 
\begin{equation}
f = \lambda + \lambda^2 R^2 
\end{equation}
is the scalar Eddington factor, $\tensor{I}$ is the identity tensor of
rank 2, and $\hat{\vec{n}} = \nabla E_{\nu} / |\nabla E_{\nu}|$.
There are other choices for the scalar Eddington factor.  For example,
\citet{Kershaw76} has suggested
\begin{equation}
  f = \frac{1}{3} + \frac{2}{3} \lambda^2 R^2.
\end{equation}
Note that in the optically thick limit, $f \rightarrow 1/3$, and in the
optically thin limit, $f \rightarrow 1$. 

Using the FLD approximation and the approximate closure, the equations
of the radiation hydrodynamics now become
\begin{align}
\frac{\partial \rho}{\partial t} + \nabla \cdot (\rho \vec{u}) = { } & 0, \label{eq:fdrhd-rho}\\
\frac{\partial (\rho \vec{u})}{\partial t} + \nabla \cdot (\rho \vec{u}
  \vec{u}) + \nabla p = { } & -\int_0^{\infty} \lambda \nabla E_{\nu}
  \mathrm{d}\nu + \vec{F}_G, \\ 
\frac{\partial (\rho E)}{\partial t} + \nabla \cdot (\rho E \vec{u} + p
  \vec{u}) 
  = { } & -\vec{u} \cdot \int_0^{\infty} \lambda \nabla E_{\nu} \mathrm{d}\nu +
  \int_0^{\infty} (c\kappa E_{\nu} - 4 \pi \eta)  \mathrm{d}\nu
  +\vec{u} \cdot \vec{F}_G, \\
\frac{\partial (\rho Y_e)}{\partial t} + \nabla \cdot (\rho Y_e \vec{u}) 
 = { } &
\int_0^{\infty} \xi (c\kappa E_{\nu} - 4\pi\eta) \mathrm{d}\nu \\
\frac{\partial E_{\nu}}{\partial t} 
+ \nabla \cdot (E_{\nu} \vec{u})
= { } & \nabla \cdot \left(\frac{c\lambda}{\chi} \nabla E_{\nu}
\right) - (c\kappa E_{\nu} - 4\pi\eta) \nonumber \\
{ } & + \frac{\partial}{\partial
  \ln{\nu}} \left(\frac{1-f}{2} E_\nu \nabla \cdot \vec{u} +
  \frac{3f-1}{2} E_\nu \hat{\vec{n}}\hat{\vec{n}} : 
  \nabla \vec{u}\right) . \label{eq:fdrhd-Enu} 
\end{align}

\subsection{Multigroup Radiation Hydrodynamics Equations}
\label{sec:mgrhd}

To numerically solve Equations
(\ref{eq:fdrhd-rho})--(\ref{eq:fdrhd-Enu}), we first discretize the
system in frequency space by dividing the radiation into $N$ energy
groups.  We define the radiation energy density in each group as
\begin{equation}
  E_g = \int_{\nu_{g-1/2}}^{\nu_{g+1/2}} E_\nu \mathrm{d}\nu, \label{eq:Egdef}
\end{equation}
where $\nu_{g-1/2}$ and $\nu_{g+1/2}$ are the frequency at the lower
and upper boundaries, respectively, for the $g$-th group.  For
clarity, $\int_{\nu_{g-1/2}}^{\nu_{g+1/2}} \mathrm{d}\nu$ is denoted
by $\int_g\mathrm{d}\nu$ hereafter.

In the multigroup methods, various group mean coefficients are
introduced to the system.  For example, the Rosseland mean interaction
coefficient is defined as
\begin{equation}
  \chi_g^{-1} = \frac{\int_g \chi^{-1} (\partial B_\nu/\partial T)
    \mathrm{d}\nu}{\int_g (\partial B_\nu/\partial T) \mathrm{d}\nu},
\end{equation}
where $B_\nu = \eta / \kappa$.  In this paper, we will simply assume
that
\begin{align}
  \kappa_g = { } & \kappa(\nu_g) , \label{eq:kappa-g} \\
  \chi_g = { } & \chi(\nu_g). \label{eq:chi-g}
\end{align}
where $\nu_g$ is the representative frequency for the $g$-th group.
Our simple treatment of group mean coefficients is sufficient for the
transport of continuum radiation.  In particular, it is sufficient for
neutrino transport in the core-collapse supernova problem because the
monochromatic opacities and emissivities of neutrinos are smooth
functions of neutrino frequency.  It should, however, be emphasized
that our algorithm for radiation hydrodynamics does not rely on the
simple treatment of group mean coefficients used in this paper.  We
also assume that the flux limiter is approximately independent of
frequency within each group, and denote the flux limiter and the
scalar Eddington factor for the $g$-th group as $\lambda_g$ and $f_g$,
respectively.  As for the emissivity, we define
\begin{equation}
  j_g = \frac{4\pi}{c}\eta(\nu_g) \Delta \nu_g,
\end{equation}
where $\Delta \nu_g = \nu_{g+1/2} - \nu_{g-1/2}$ is the group width.
However, in the case of photon radiation, we have adopted the
polylogarithm function based method of \citet{Clark87} for computing
the group-integrated Planck function.  Thus, assuming local
thermodynamic equilibrium, we have for photon radiation
\begin{equation}
  j_g = \frac{4\pi}{c}\kappa_g B_g, 
\end{equation}
where $B_g = \int_g B_\nu \mathrm{d}\nu$.

Using these definitions and approximations, we obtain the multigroup
radiation hydrodynamics equations
\begin{align}
  \frac{\partial \rho}{\partial t} + \nabla \cdot (\rho \vec{u})
  = { } & 0, \label{eq:mgrhd-rho} \\
    \frac{\partial (\rho \vec{u})}{\partial t} + \nabla \cdot (\rho \vec{u}
  \vec{u}) + \nabla p + \sum_{g}
  \lambda_g \nabla E_g = { } &  \vec{F}_G, \label{eq:mgrhd-rhou} \\
  \frac{\partial (\rho E)}{\partial t} + \nabla \cdot (\rho E \vec{u} + p
  \vec{u}) + \vec{u} \cdot \sum_{g}\lambda_g
  \nabla E_g  = { } & \sum_{g} c
  (\kappa_gE_{g}-j_g) + \vec{u}\cdot\vec{F}_G, \label{eq:mgrhd-rhoE} \\
  \frac{\partial (\rho Y_e)}{\partial t} + \nabla \cdot (\rho Y_e \vec{u})
  = { } & \sum_{g} c \xi_g (\kappa_gE_{g}-j_g)
  , \label{eq:mgrhd-Ye} \\
\frac{\partial E_g}{\partial t} + \nabla \cdot 
  \left(\frac{3-f_g}{2} E_g \vec{u}\right) - \vec{u} \cdot \nabla
  \left(\frac{1-f_g}{2} E_g\right)  = { } & 
    - c (\kappa_gE_{g}-j_g)
  + \nabla \cdot \left(\frac{c\lambda_g}{\chi_g} \nabla E_g \right)
   \label{eq:mgrhd-Eg} \\
  + \int_g \frac{\partial}{\partial \nu} \Bigg{[}\left(\frac{1-f}{2}
    \nabla \cdot \vec{u} + \frac{3f-1}{2} \hat{\vec{n}}\hat{\vec{n}}
    : \nabla \vec{u} \right) & \nu E_\nu \Bigg{]} \mathrm{d}\nu -
    \frac{3f_g-1}{2} E_g \hat{\vec{n}}\hat{\vec{n}} : \nabla \vec{u}, \nonumber
\end{align}
where $g = 1, 2, \ldots, N$, $\xi_g = \xi(\nu_g)$, $\sum_g$ denotes
$\sum_{g=1}^N$.

\subsection{Mathematical Characteristics of the Hyperbolic Subsystem}
\label{sec:hyper}

The system of multigroup radiation hydrodynamics
(Equations \ref{eq:mgrhd-rho}--\ref{eq:mgrhd-Eg}), similar to the system of
gray radiation hydrodynamics we have studied in Paper II, has a
hyperbolic subsystem,
\begin{align}
  \frac{\partial \rho}{\partial t} + \nabla \cdot (\rho \vec{u})
  = { } & 0, \label{eq:hyper-rho} \\
    \frac{\partial (\rho \vec{u})}{\partial t} + \nabla \cdot (\rho \vec{u}
  \vec{u}) + \nabla p + \sum_{g}
  \lambda_g \nabla E_g = { } & \vec{F}_G , \label{eq:hyper-rhou} \\
  \frac{\partial (\rho E)}{\partial t} + \nabla \cdot (\rho E \vec{u} + p
  \vec{u}) +\vec{u} \cdot \sum_{g} \lambda_g
   \nabla E_g = { } & \vec{u} \cdot \vec{F}_G , \label{eq:hyper-rhoE} \\
  \frac{\partial (\rho Y_e)}{\partial t} + \nabla \cdot (\rho Y_e \vec{u})
  = { } & 0, \label{eq:hyper-Ye} \\
    \frac{\partial E_g}{\partial t} + \nabla \cdot 
  \left(\frac{3-f_g}{2} E_g \vec{u}\right) - \vec{u} \cdot
  \nabla \left( \frac{1-f_g}{2} E_g\right)  = { } & 0 ,
  \label{eq:hyper-Eg}
\end{align}
where $g = 1, 2, \ldots, N$.  Since $Y_e$ is just an auxiliary
variable that does not alter the characteristic wave speeds of the
hyperbolic subsystem, we will neglect it in the
remainder of this subsection for simplicity.  Gravitational force terms
are source terms, and thus they are not included in the following
analysis.  The term $-[(3f_g-1)/2]E_g \hat{\vec{n}}\hat{\vec{n}} :
\nabla \vec{u}$ in Equation~(\ref{eq:mgrhd-Eg}) is not included in the
analysis of the hyperbolic subsystem here even though it becomes
comparable to the term $\nabla \cdot \{[(3-f_g)/2] E_g \vec{u}\}$ in
Equation~(\ref{eq:hyper-Eg}) in the streaming limit where $f_g
\rightarrow 1$.  The first reason for not including the term is that
the main purpose of including radiation in the hyperbolic subsystem is
to improve the accuracy in the optically thick limit where the
radiation can affect the hydrodynamics significantly.  The term
$-[(3f_g-1)/2]E_g \hat{\vec{n}}\hat{\vec{n}} : \nabla \vec{u}$ is not
significant in the optically thick limit where $f_g \rightarrow 1/3$.
The second reason is that in the streaming limit, where $\lambda
\rightarrow 0$, radiation force has little impact on the motion of
matter.  Thus, radiation is essentially decoupled from the
hydrodynamics of matter in the streaming limit.  Moreover, in the
streaming limit the diffusion term $\nabla \cdot [(c\lambda_g/\chi_g)
\nabla E_g]$ (Equation~\ref{eq:mgrhd-Eg}) will dominate the term
$-[(3f_g-1)/2]E_g \hat{\vec{n}}\hat{\vec{n}} : \nabla \vec{u}$ and all
the terms in Equation~(\ref{eq:hyper-Eg}).  Therefore, the term
$-[(3f_g-1)/2]E_g \hat{\vec{n}}\hat{\vec{n}} : \nabla \vec{u}$ is
insignificant in both the diffusion limit and the streaming limit.

In CASTRO, we solve the hyperbolic subsystem with a Godunov method,
which utilizes a characteristic-based Riemann solver (see
\S~\ref{sec:explicit} for details).  The Godunov method requires that
we analyze the mathematical characteristics of the hyperbolic
subsystem.  For simplicity, let us consider the system in one
dimension, which can be written in terms of primitive variables as,
\begin{equation}
  \frac{\partial{Q}}{\partial{t}} + A \frac{\partial{Q}}{\partial{x}}
  = 0,
\end{equation}
where the primitive variables are
\begin{equation}
  Q = \left( \begin{array}{c}
              \rho \\
              u \\
              p \\
              E_1 \\
              E_2 \\
              \vdots \\
              E_N \end{array}\right),
\end{equation}
and the Jacobian matrix is
\begin{equation}
  A = \left( \begin{array}{ccccccc}
      u & \rho & 0 & 0 & 0 & \ldots & 0 \\
      0 & u    & {1}/{\rho} & {\lambda_1}/{\rho} &
        {\lambda_2}/{\rho} & \ldots & {\lambda_N}/{\rho} \\
      0 & \gamma p & u & 0 & 0 & \ldots & 0 \\
      0 & E_1 (3-f_1)/2 & 0 & u & 0 & \ldots & 0 \\ 
      0 & E_2 (3-f_2)/2 & 0 & 0 & u & \ldots & 0 \\
      \vdots & \vdots & \vdots & \vdots & \vdots & \ddots & \vdots \\
      0 & E_N (3-f_N)/2 & 0 & 0 & 0 & \ldots & u 
      \end{array} \right),
\end{equation} 
where $\gamma$ is the adiabatic index of the matter.  This system is
hyperbolic because the Jacobian matrix is diagonalizable with $N+3$
real eigenvalues,
\begin{equation}
  u-c_s, \ u+c_s, \ u, \ u, \ u, \ldots, \ u.
\end{equation}
Here
\begin{equation}
c_s = \sqrt{\gamma \frac{p}{\rho} + \sum_{g}\left(\frac{3-f_g}{2}\right) \left(\frac{\lambda_g
    E_g}{\rho}\right)} = \sqrt{c_m^2 + \sum_{g} c_g^2} \label{eq:cs}
\end{equation}
is the radiation modified sound speed, where $c_m = \sqrt{\gamma p /
  \rho}$ is the sound speed without radiation and $c_g =
\sqrt{[(3-f_g)/2] (\lambda_g E_g / \rho)}$ is the contribution to the
sound speed from each group.  The corresponding right eigenvectors
are,
\begin{equation}
\left(\begin{array}{c} 1 \\ -c_s/\rho \\ \gamma p/\rho \\
    c_1^2/\lambda_1 \\ c_2^2/\lambda_2 
    \\ \vdots 
    \\ c_N^2/\lambda_N \end{array}\right), \ 
\left(\begin{array}{c} 1 \\ c_s/\rho \\ \gamma p/\rho \\
    c_1^2/\lambda_1 \\ c_2^2/\lambda_2 
    \\ \vdots 
    \\ c_N^2/\lambda_N \end{array}\right), \ 
\left(\begin{array}{c} 1 \\ 0 \\ 0 \\ 0 \\ 0 
    \\ \vdots
  \\ 0 \end{array} \right), \ 
\left(\begin{array}{c} 0 \\ 0 \\ -\lambda_1 \\ 1 \\ 0
    \\ \vdots
    \\ 0\end{array} \right), \ 
\left(\begin{array}{c} 0 \\ 0 \\ -\lambda_2 \\ 0 \\ 1
    \\ \vdots
    \\ 0\end{array} \right), \ \ldots,
\left(\begin{array}{c} 0 \\ 0 \\ -\lambda_N \\ 0 \\ 0
    \\ \vdots
    \\ 1\end{array} \right), \ 
\end{equation}
and the corresponding left eigenvectors are,
\begin{equation}
  \begin{array}{rrrrrrr}
  (0,& -\rho/2c_s,& 1/2{c_s}^2,& \lambda_1/2 {c_s}^2,& \lambda_2/2
        {c_s}^2,& \ldots ,& \lambda_N/2 {c_s}^2), \\
  (0,& \rho/2c_s,& 1/2{c_s}^2,& \lambda_1/2 {c_s}^2,& \lambda_2/2
        {c_s}^2,& \ldots ,& \lambda_N/2 {c_s}^2), \\
  (1,&             0,& -1 / {c_s}^{2},& -\lambda_1 / {c_s}^2,& -\lambda_2
       / {c_s}^2,& \ldots ,
      & -\lambda_N / {c_s}^2 ), \\
  (0,&             0,& -\omega_1/\lambda_1,& 1- \omega_1,&
  -\omega_1\lambda_2/\lambda_1, & \ldots ,& -\omega_1 \lambda_N/\lambda_1), \\
  (0,&             0,& -\omega_2/\lambda_2,& -\omega_2\lambda_1/\lambda_2,&
  1-\omega_2,& \ldots ,& -\omega_2\lambda_N/\lambda_2), \\ 
     & &  & \vdots & & &, \\
  (0,&             0,& -\omega_N/\lambda_N,& - \omega_N\lambda_1/\lambda_N,&
      -\omega_N\lambda_2/\lambda_N,& \ldots ,& 1-\omega_N),
  \end{array} 
\end{equation}
where
\begin{equation}
  \omega_g = \frac{c_g^2}{c_s^2}.
\end{equation}
These $N+3$ eigenvectors define the characteristic fields for the
one-dimensional system.  By computing the product of the right
eigenvectors and the gradients of their corresponding eigenvalues
\citep{LeVeque}, we find that the first and second fields are
genuinely nonlinear corresponding to either a shock wave or a
rarefaction wave.  The rest of the fields are linearly degenerate
corresponding to a contact discontinuity in either density or
radiation energy density of one group moving at a speed of $u$.  Note
that there is also a jump in fluid pressure such that the total
pressure, $p_{\mathrm{tot}} = p + \sum_{g}\lambda_g E_g$, is constant
across the contact discontinuity.  Obviously, in three dimensions,
there are two additional contact discontinuities for transverse
velocities just like the case of pure hydrodynamics.

\subsection{Advection in Frequency Space}
\label{sec:fspace}

The integrals in Equation~(\ref{eq:mgrhd-Eg}) form the following
system,
\begin{equation}
 \frac{\partial E_g}{\partial t} = \int_g \frac{\partial}{\partial
   \nu}  \left[\left(\frac{1-f}{2} 
    \nabla \cdot \vec{u} + \frac{3f-1}{2} \hat{\vec{n}}\hat{\vec{n}}
    : \nabla \vec{u} \right) \nu E_\nu\right] \mathrm{d}\nu,
 \label{eq:fspace}
\end{equation}
which can also be written in the conservation law form
\begin{equation}
  \frac{\partial q_1}{\partial t} + \frac{\partial (a_qq_1)}{\partial
    \ln{\nu}} = 0, \label{eq:fsp-e}
\end{equation}
where $q_1 = \nu E_\nu$ and $a_q = -[(1-f)/2] (\nabla \cdot \vec{u})
- [(3f-1)/2] \hat{\vec{n}}\hat{\vec{n}} : \nabla \vec{u}$.  Here we
have used Equation~(\ref{eq:Egdef}). 
Equation~(\ref{eq:fsp-e}) describes the advection of $q_1$ in
logarithmic frequency space with an advection speed of $a_q$.  Because
$\int q_1 \mathrm{d}\ln{\nu}$ is the integrated radiation energy, the
system conserves energy.

Various terms in Equation~(\ref{eq:mgrhd-Eg}) have been split between
the hyperbolic subsystem (\S~\ref{sec:hyper}) and the system for the
advection in the frequency space (\S~\ref{sec:fspace}).  There is
however another way to look at Equation~(\ref{eq:mgrhd-Eg}).  The
evolution of the radiation energy density (Equation~\ref{eq:mgrhd-Eg})
without the diffusion term $\nabla \cdot [(c\lambda_g/\chi_g) \nabla
E_g]$ and the source-sink term $-c(\kappa_g E_g - j_g)$ can also be
split as
\begin{align}
  \frac{\partial E_g}{\partial t}  = { } &
   -\nabla \cdot (E_g \vec{u}), \label{eq:Eg2} \\
\frac{\partial E_\nu}{\partial t} = { } &
  \frac{\partial}{\partial
  \ln{\nu}} \left[ \left(\frac{1-f}{2} \nabla \cdot \vec{u} +
  \frac{3f-1}{2} \hat{\vec{n}}\hat{\vec{n}} : 
  \nabla \vec{u}\right) E_\nu \right]. \label{eq:fspace2}
\end{align}
Equation~(\ref{eq:fspace2}) also represents an advection in
logarithmic frequency space and can be written in the form
\begin{equation}
  \frac{\partial q_2}{\partial t} + \frac{\partial (a_q q_2)}{\partial
    \ln{\nu}} = 0, \label{eq:fsp-n}
\end{equation}
where $q_2 = E_\nu$. Here the conserved quantity is the number of
radiation particles (i.e., $\int E_\nu/\nu\, \mathrm{d} \nu$).  This
new way of splitting will be further discussed later in
Section~\ref{sec:explicit}.

\subsection{Multigroup Radiation Diffusion} 
\label{sec:para}

The parabolic part of the system consists of the radiation diffusion
and source-sink terms, which were omitted from the discussion of the
hyperbolic subsystem (\S~\ref{sec:hyper}) and advection in
frequency space (\S~\ref{sec:fspace}),
\begin{align}
\frac{\partial (\rho e)}{\partial t} = { } & \sum_{g} c
  (\kappa_gE_{g}-j_g), \label{eq:drhoe-i} \\
\frac{\partial (\rho Y_e)}{\partial t} = { } & \sum_{g} c \xi_g
  (\kappa_gE_{g}-j_g), \label{eq:dYe-i}\\
\frac{\partial E_g}{\partial t} = { } & -c (\kappa_gE_{g}-j_g)
  + \nabla \cdot \left(\frac{c\lambda_g}{\chi_g} \nabla E_g \right),
   \label{eq:dEg-i}
\end{align}
where $e$ is the specific internal energy of the fluid, and $g = 1, 2,
\ldots, N$.  The term $c (\kappa_{g} E_{g} - j_g)$ represents the
energy exchange in the comoving frame between the matter and the
$g$-th radiation group through absorption and emission of
radiation.  Note that the terms on the right-hand side (RHS) of
Equations~(\ref{eq:drhoe-i}), (\ref{eq:dYe-i}), and (\ref{eq:dEg-i})
all contain the speed of light $c$.  Thus these are order $O(1)$
terms in $v/c$.  An implicit treatment for these equations is usually necessary
because of their stiffness.  For photon radiation,
Equation~(\ref{eq:dYe-i}) is not needed.

\section{Numerical Methods}
\label{eq:method}

CASTRO uses a nested hierarchy of logically-rectangular,
variable-sized grids with simultaneous refinement in both space and
time.  The AMR algorithms for hyperbolic and radiation diffusion
equations in CASTRO have been described in detail in Papers I and II,
respectively.  The multigroup solver does not require any changes in
the AMR algorithms.  Thus, we will not describe them again in this
paper.

In this section, we describe the single-level integration scheme.  For
each step at a single level of refinement, the state is first evolved
using an explicit method for the hyperbolic subsystem
(Equations~\ref{eq:hyper-rho}--\ref{eq:hyper-Eg}) and advection in
frequency space (\S~\ref{sec:explicit}).  Then an implicit update for
radiation diffusion and source-sink terms is performed
(\S~\ref{sec:implicit}).  In \S~\ref{sec:accel}, we describe two
acceleration schemes for speeding up the convergence rate of the
implicit solver for multigroup radiation diffusion equations.

\subsection{Explicit Solver for Hyperbolic subsystem and Advection in
  Frequency Space}
\label{sec:explicit}

The hyperbolic subsystem (\S~\ref{sec:hyper}) is treated explicitly.
Our Godunov algorithm for the hyperbolic subsystem of the multigroup
radiation hydrodynamics is an extension of the hyperbolic solver for
the gray radiation hydrodynamics presented in Paper II , which in turn
is based on the hydrodynamics algorithm presented in Paper I of this
series.  We refer the reader to Papers I and II for a detailed
description of the integration scheme, which supports a general
equation of state, self-gravity, nuclear reactions, the advection of
nuclear species and electron fraction.  Here we will only present a
brief overview of the Godunov scheme.

The time evolution of the hyperbolic subsystem can be written in the
form
\begin{equation}
  \frac{\partial \vec{U}}{\partial{t}} = - \nabla \cdot \vec{F},
\end{equation}
where $\vec{U} = (\rho, \rho \vec{u}, \rho E, E_1, E_2, \ldots,
E_N)^{T}$ denotes the conserved variables, and $\vec{F}$ is their flux.  The
conserved variables are defined at cell centers.  We predict the
primitive variables, including $\rho$, $\vec{u}$, $p$, $\rho e$,
$E_g$, where $g = 1, 2, \ldots, N$, from cell centers at time $t^n$ to
edges at time $t^{n+1/2}$ and use an approximate Riemann solver to
construct fluxes, $\vec{F}^{n+1/2}$, on cell faces.  This algorithm is
formally second order in both space and time.  The time step is
computed using the standard Courant-Friedrichs-Lewy (CFL) condition
for explicit methods, and the sound speed used in the computation is
now the radiation modified sound speed $c_s$ (Equation~\ref{eq:cs}).
The time step is also limited by maximally allowed changes in
temperature and electron fraction in the implicit solver.  Additional
constraints for the time step are applied if additional physics, such
as burning, is included.  CASTRO solves the hyperbolic subsystem of
radiation hydrodynamics with an unsplit piecewise parabolic method
(PPM) with characteristic tracing and full corner coupling
\citep{unsplitPPM}.  An approximate Riemann solver based on
the Riemann solver of \citet{bellcolellatrangenstein} and
\citet{ColellaGF97}, that utilizes the eigenvectors of the system, is
used to compute the Godunov states and fluxes at the cell interface.

Advection in frequency space (\S~\ref{sec:fspace}) can also be
treated explicitly because its time scale is comparable to that of the
hyperbolic subsystem.  One approach is to adopt an operator-splitting
method and to evolve Equation~(\ref{eq:fsp-e}) with a Godunov method.
This is the approach taken by \citet{crash} in the CRASH code for
photon radiation.  The term $-[(3f_g-1)/2]E_\nu
\hat{\vec{n}}\hat{\vec{n}} : \nabla \vec{u}$ can be treated as a
source term of Equation~(\ref{eq:fsp-e}).  The success of this
approach for a number of test problems has been demonstrated by
\citet{crash}.  We have also found that this approach can produce
satisfactory results for all the photon test problems in
Section~\ref{sec:tests}.  However, we have found that erroneous
results were obtained for core-collapse supernova simulations
(\S~\ref{test-n}), probably due to operator-splitting errors.  One
difficulty in core-collapse supernova simulations is that the
divergence of velocity can be significant because there is a strong
shock and the typical velocity in front of the shock is $\sim
0.1$--$0.2\, c$.  In the case of core-collapse supernovae, the
negative divergence of velocity will shift neutrinos towards high
frequency (Equation~\ref{eq:fsp-e}).  The stronger conservative
scattering at higher frequency can further amplify the neutrino energy
density due to the work term $\vec{u} \cdot \nabla \{[(1-f_g)/2]E_g\}$
(Equation~\ref{eq:hyper-Eg}).  These two processes can combine to
cause unphysical results due to splitting errors.  However, the origin
of the various terms involved in these two processes is the term
$\partial(\tensor{P}_\nu : \nabla \vec{u})/\partial \ln{\nu}$ in
Equation~(\ref{eq:dEnudt}).  To avoid operator-splitting errors, it is
therefore desirable to treat the term $\partial(\tensor{P}_\nu :
\nabla \vec{u})/\partial \ln{\nu}$ without splitting it.  On the other
hand, the term can also make significant contribution to the
hyperbolic subsystem in the optically thick limit, and we do not wish
to sacrifice accuracy of the hyperbolic subsystem by taking the
traditional approach of leaving out radiation from the Riemann solver
for the hydrodynamics.

To avoid the operator-splitting without sacrificing the accuracy of
the hyperbolic subsystem, we have adopted a new approach based on
splitting the radiation energy density equation into
Equations~(\ref{eq:Eg2}) and (\ref{eq:fspace2}).  Note that the
radiation diffusion and source-sink terms will be treated separately by
an implicit solver (\S~\ref{sec:implicit}).  Our new approach still
utilizes the Godunov scheme for the hyperbolic subsystem governed by
Equations~(\ref{eq:hyper-rho})--(\ref{eq:hyper-Eg}).  After the
Godunov states and fluxes at the cell interfaces are obtained, all
variables in the hyperbolic system except the radiation energy density
are updated as usual.  For the radiation energy density, two steps are
taken.  The radiation energy density is first updated according to
Equation~(\ref{eq:Eg2}) with its RHS computed using the Godunov
states.  Then another Godunov solver is used to evolve
Equation~(\ref{eq:fsp-n}), which is equivalent to
Equation~(\ref{eq:fspace2}).  The conversion between $E_g$ and $E_\nu$
is performed according to $E_g = E_\nu(\nu_g) \nu_g \Delta
\ln{\nu_g}$. The Godunov states at time $t^{n+1/2}$ of the first
Godunov solver are used in computing $\nabla \cdot \vec{u}$ and
$\nabla \vec{u}$ to obtain the wave speed $a_q$ in the frequency space
advection.  The second Godunov scheme for the frequency space
advection is a standard approach based on the method of lines.  For
time integration we use the third-order total variation diminish
Runge-Kutta scheme of \citet{ShuOsher88}.  We use the piecewise linear
method with a monotonized central slope limiter \citep{MC} to
reconstruct $q_2$ and $a_q$ in order to achieve high order in
(logarithmic frequency) space.  The reconstructed variables are used
to compute the HLLE flux \citep{HLL, HLLE}.  The frequency space
advection has its own CFL condition for stability.  Since CASTRO uses
the hydrodynamic CFL condition with other additional constraints for
the time steps, if necessary, subcycling in time is employed in order
to satisfy the frequency space CFL condition.

\subsection{Implicit Solver for Radiation Diffusion and Source-Sink Terms}
\label{sec:implicit}

The implicit solver evolves the radiation and matter according to
equations (\ref{eq:drhoe-i}), (\ref{eq:dYe-i}), and (\ref{eq:dEg-i})
with the results of the explicit update (\S~\ref{sec:explicit}) as
the initial conditions.  Our scheme is based on a first-order
backward Euler method.

\subsubsection{Outer Newton Iteration}
\label{sec:outerIter}

We solve the parabolic subsystem (Equations~\ref{eq:drhoe-i},
\ref{eq:dYe-i}, and \ref{eq:dEg-i}) iteratively via Newton's method.
We define
\begin{align}
  F_e = { } & \rho e - \rho e^{-} - \Delta t \sum_g c (\kappa_g E_g -
  j_g) , \\
  F_Y = { } & \rho Y_e - \rho Y_e^{-} - \Delta t \sum_g c \xi_g (\kappa_g
  E_g - j_g) , \\
  F_g = { } & E_g - E_g^{-} - \Delta t\, \nabla \cdot
  \left(\frac{c\lambda_g}{\chi_g} \nabla E_g \right)  
  + \Delta t\, c (\kappa_g E_g - j_g)  ,
\end{align}
where the $-$ superscript denotes the state following the explicit
update, and $g = 1, 2, \ldots, N$.  The desired solution is for $F_g$,
$F_e$ and $F_Y$ to all be zero.  We approach the solution by Newton
iteration, and in each iteration step we solve the following system
\begin{equation}
  \left[\begin{array}{ccc}
      ({\partial F_e}/{\partial T})^{(k)}
      & ({\partial F_e}/{\partial Y_e})^{(k)}
      & ({\partial F_e}/{\partial E_g})^{(k)} \\[3pt]
      ({\partial F_Y}/{\partial T})^{(k)}
      & ({\partial F_Y}/{\partial Y_e})^{(k)}
      & ({\partial F_Y}/{\partial E_g})^{(k)} \\[3pt]
      ({\partial F_g}/{\partial T})^{(k)}
      & ({\partial F_g}/{\partial Y_e})^{(k)}
      & ({\partial F_g}/{\partial E_g})^{(k)}
      \end{array} \right]
  \left[\begin{array}{c}
      \delta T^{(k+1)}\\
      \delta Y_e^{(k+1)}\\
      \delta E_g^{(k+1)}
      \end{array}\right]
= \left[\begin{array}{c}
      - F_e^{(k)}\\
      - F_Y^{(k)}\\
      - F_g^{(k)}
      \end{array}\right]. \label{eq:newton}
\end{equation}
Here $\delta T^{(k+1)} = T^{(k+1)} - T^{(k)}$, $\delta Y_e^{(k+1)} =
Y_e^{(k+1)} - Y_e^{(k)}$ and $\delta E_g^{(k+1)} = E_g^{(k+1)} -
E_g^{(k)}$, where the $(k)$ superscript denotes the stage of the
Newton iteration.  The exact Jacobian matrix in equation
(\ref{eq:newton}) is very complicated.  For simplicity, we neglect the
derivatives of the diffusion coefficient $c\lambda_g/\chi_g$.  This is
a common practice in radiation transport, and does not appear to
significantly degrade the convergence rate.  With the approximate
Jacobian matrix, we can eliminate $\delta T$ and $\delta Y_e$ from
equation (\ref{eq:newton}) and obtain the following equation for
$E_g^{(k+1)}$,
\begin{equation} \begin{split}
  \left(c\kappa_g + \frac{1}{\Delta t}\right) E_g^{(k+1)} & - \nabla \cdot \left(
      \frac{c\lambda_g}{\chi_g} \nabla E_g^{(k+1)} \right) = c j_g +
      \frac{E_g^{-}}{\Delta t} \\ 
   & + H_g
  \left[c\sum_{g^{\prime}}\left(\kappa_{g^{\prime}} E_{g^{\prime}}^{(k+1)} -
    j_{g^{\prime}} \right) - \frac{1}{\Delta t} (\rho e^{(k)} - \rho e^{-}) \right] \\ 
  & +  \Theta_g
  \left[c \sum_{g^{\prime}}\xi_{g^{\prime}}\left(\kappa_{g^{\prime}}
      E_{g^{\prime}}^{(k+1)} - j_{g^{\prime}} \right) - \frac{1}{\Delta t}
    (\rho Y_e^{(k)} - \rho Y_e^{-}) \right]. \label{eq:MGdiff0} 
\end{split}\end{equation}
To reduce clutter, we have dropped the $(k)$ superscript for
$\lambda_g$, $\kappa_g$, $\chi_g$, and $j_g$.  The newly introduced
variables in equation (\ref{eq:MGdiff0}) are given by
\begin{align}
H_g = { } & \left(\frac{\partial j_g}{\partial T} - \frac{\partial
    \kappa_g}{\partial T} E_g^{(k)} \right) \eta_T -
    \left(\frac{\partial j_g}{\partial Y_e} - \frac{\partial
    \kappa_g}{\partial Y_e} E_g^{(k)}\right)  \eta_Y ,\\  
\Theta_g = { } & \left(\frac{\partial j_g}{\partial Y_e} - \frac{\partial
    \kappa_g}{\partial Y_e} E_g^{(k)}\right) \theta_Y - 
    \left(\frac{\partial j_g}{\partial T} - \frac{\partial
    \kappa_g}{\partial T} E_g^{(k)} \right) \theta_T , 
\end{align}
where
\begin{align}
  \eta_T = { } & \frac{c\Delta t}{\Omega} \left[\rho + c \Delta t
    \sum_g \xi_g \left(\frac{\partial j_g}{\partial Y_e} - \frac{\partial
    \kappa_g}{\partial Y_e} E_g^{(k)}\right) \right]  , \\
  \eta_Y = { } & \frac{c\Delta t}{\Omega} \left[c \Delta t \sum_g
    \xi_g \left(\frac{\partial j_g}{\partial T} - \frac{\partial
    \kappa_g}{\partial T} E_g^{(k)} \right) \right] , \\
  \theta_T = { } & \frac{c\Delta t}{\Omega} \left[ \rho \frac{\partial
      e}{\partial Y_e} + c \Delta t \sum_g 
      \left(\frac{\partial j_g}{\partial Y_e} - \frac{\partial
      \kappa_g}{\partial Y_e} E_g^{(k)}\right) \right] , \\ 
  \theta_Y = { } & \frac{c\Delta t}{\Omega} \left[\rho \frac{\partial
      e}{\partial T} + c\Delta t \sum_g 
      \left(\frac{\partial j_g}{\partial T} - \frac{\partial
    \kappa_g}{\partial T} E_g^{(k)} \right) \right] , \\
  \Omega = { } & \left[\rho \frac{\partial
      e}{\partial T} + c\Delta t \sum_g 
      \left(\frac{\partial j_g}{\partial T} - \frac{\partial
    \kappa_g}{\partial T} E_g^{(k)} \right) \right]
      \left[\rho + c \Delta t
    \sum_g \xi_g \left(\frac{\partial j_g}{\partial Y_e} - \frac{\partial
    \kappa_g}{\partial Y_e} E_g^{(k)}\right) \right]
  \nonumber \\
   & - \left[ \rho \frac{\partial
      e}{\partial Y_e} + c \Delta t \sum_g 
      \left(\frac{\partial j_g}{\partial Y_e} - \frac{\partial
      \kappa_g}{\partial Y_e} E_g^{(k)}\right) \right]
     \left[c \Delta t \sum_g
    \xi_g \left(\frac{\partial j_g}{\partial T} - \frac{\partial
    \kappa_g}{\partial T} E_g^{(k)} \right) \right].
\end{align}
Here all the derivatives are computed from the results of the previous
Newton iteration (i.e., the $(k)$ state).

\subsubsection{Inner Iteration}
\label{sec:innerIter}

We now present our algorithm for solving equation (\ref{eq:MGdiff0}),
which is actually a set of $N$ equations coupled through the summation
terms on the RHS of equation (\ref{eq:MGdiff0}).  The system is
usually too large to be solved directly.  Instead, we solve it by
decoupling the groups and utilizing an iterative method.  Note that
the iteration here for solving equation (\ref{eq:MGdiff0}) is
different from the Newton iteration for solving the entire parabolic
subsystem.  The inner iteration here is embedded inside the outer
Newton iteration step.  In our algorithm, we avoid the coupling by
replacing the radiation energy density in the coupling terms with its
state from the previous inner iteration.  Thus, each step of the inner
iteration solves
\begin{equation} \begin{split}
  \left(c\kappa_g + \frac{1}{\Delta t}\right) E_g^{\ell+1} & - \nabla \cdot \left(
      \frac{c\lambda_g}{\chi_g} \nabla E_g^{\ell+1} \right) = c j_g +
      \frac{E_g^{-}}{\Delta t} \\ 
   & + H_g
  \left[c\sum_{g^{\prime}}\left(\kappa_{g^{\prime}} E_{g^{\prime}}^{\ell} -
    j_{g^{\prime}} \right) - \frac{1}{\Delta t} (\rho e^{(k)} - \rho e^{-}) \right] \\ 
  & +  \Theta_g
  \left[c \sum_{g^{\prime}}\xi_{g^{\prime}}\left(\kappa_{g^{\prime}}
      E_{g^{\prime}}^{\ell} - j_{g^{\prime}} \right) - \frac{1}{\Delta t}
    (\rho Y_e^{(k)} - \rho Y_e^{-}) \right]. \label{eq:MGdiff1} 
\end{split}\end{equation}
where $\ell$ is the inner iteration index.  Here, we have dropped the
$(k+1)$ superscript for $E_g^{\ell+1}$ and $E_g^{\ell}$ to reduce
clutter.  For the first inner iteration, the state of the previous
inner iteration is set to the result of the previous outer iteration.
The inner iteration for equation (\ref{eq:MGdiff1}) is stopped when
the maximum of $\sum_g |(E_g^{\ell+1} -E_g^\ell)| /
\sum_gE_g^{\ell+1}$ on the computational domain is below a preset
tolerance (e.g., $10^{-6}$).

CASTRO can solve equations in the following canonical form (Paper II)
\begin{equation}
A E_g^{\ell+1}
- \sum_i \frac{\partial}{\partial x^i} \left(B_i \frac{\partial
      E_g^{\ell+1}}{\partial x^i} \right)
+ \sum_i \frac{\partial}{\partial x^i} \left( C_i E_g^{\ell+1} \right)
+ \sum_i D_i \frac{\partial E_g^{\ell+1}}{\partial x^i}
= \mathrm{RHS},\label{eq:canon}
\end{equation}
where $A$ are cell-centered coefficients and $B_i$, $C_i$, and $D_i$
are centered at cell faces.  We use the same approach for spatial
discretization at the AMR coarse-fine interface as in \citet{IAMR}.
For equation (\ref{eq:MGdiff1}) the $C_i$ and $D_i$ coefficients are
not used.  The same canonical form works for Cartesian, cylindrical,
and spherical coordinates so long as appropriate metric factors are
included in the coefficients and the RHS.  CASTRO uses the {\it hypre}
library \citep{hypre,hypreweb} to solve the linear system.

\subsubsection{Matter Update}
\label{sec:upm}

When the numerical solution to equation (\ref{eq:MGdiff0}) is obtained
via the inner iteration (Equation~\ref{eq:MGdiff1}), we update the
matter and continue with the next outer iteration until convergence is
achieved.  Using Equation~(\ref{eq:newton}), we can obtain
\begin{align}
  \delta T^{(k+1)} = { } & \eta_T \left[\sum_g(\kappa_gE_g^{(k+1)} - j_g) -
    \frac{\rho e^{(k)} - \rho e^-}{c\Delta t} \right] \nonumber \\ 
    & - \theta_T\left[
    \sum_g\xi_g(\kappa_gE_g^{(k+1)} - j_g) - \frac{\rho
      Y_e^{(k)} - \rho Y_e^-}{c\Delta t} \right] , \label{eq:dT} \\
  \delta Y_e^{(k+1)} = { } & -\eta_Y \left[\sum_g(\kappa_gE_g^{(k+1)} - j_g) -
    \frac{\rho e^{(k)} - \rho e^-}{c\Delta t} \right] \nonumber \\
     & + \theta_Y \left[
    \sum_g\xi_g(\kappa_gE_g^{(k+1)} - j_g) - \frac{\rho Y_e^{(k)} - \rho
      Y_e^-}{c\Delta t} \right].\label{eq:dY}
\end{align}
Thus we could update the matter temperature as $T^{(k+1)} = T^{(k)} +
\delta T^{(k+1)}$ and electron fraction as $Y_e^{(k+1)} = Y_e^{(k)} +
\delta Y_e^{(k+1)}$, and then compute the matter specific internal
energy $e^{(k+1)}$ from $T^{(k+1)}$ and $Y_e^{(k+1)}$.  However, the
total energy (radiation energy plus matter internal energy) and lepton
number would not be conserved in this type of updates.  Furthermore,
our experience with this approach is that sometimes the update gives
$T$ and $Y_e$ that are out of the bounds of opacity and EOS tables
used in our core-collapse simulations and hence causes convergence
issues.  Thus a different approach is taken in CASTRO.  To conserve
the total energy and lepton number, we update the matter as follows,
\begin{align}
  \rho e^{(k+1)} = { } &  H \rho e^{(k)} + (1-H) \rho e^- + \Theta (\rho Y_e^{(k)} -
  \rho Y_e^-) \nonumber \\
  & + c \Delta t \sum_g\left[(\kappa_gE_g^{\ell+1} - j_g) - (H
    + \Theta \xi_g) (\kappa_gE_g^{\ell} - j_g)\right] , \label{eq:uprhoe}\\ 
 \rho Y_e^{(k+1)} = { } & \bar{\Theta} \rho Y_e^{(k)} +
  (1-\bar{\Theta}) \rho Y_e^- + \bar{H} (\rho e^{(k)} - \rho
  e^-) \nonumber \\
  & + c\Delta t \sum_g\left[\xi_g(\kappa_gE_g^{\ell+1} - j_g) -
    (\bar{H} + \bar{\Theta} \xi_g) (\kappa_gE_g^{\ell} -
    j_g)\right] , \label{eq:uprhoYe}
\end{align}
where $H = \sum_g H_g$, $\Theta = \sum_g \Theta_g$, $\bar{H} = \sum_g
\xi_g H_g$, and $\bar{\Theta} = \sum_g \xi_g \Theta_g$.  Then we
compute temperature $T^{(k+1)}$ and electron fraction $Y_e^{(k+1)}$
from the newly updated $\rho e^{(k+1)}$ and $\rho Y_e^{(k+1)}$.  We
also compute the new absorption, scattering and emission coefficients
using the new temperature and electron fraction.  By updating the
matter using equations (\ref{eq:uprhoe}) and (\ref{eq:uprhoYe}), the
implicit algorithm conserves the total energy and electron lepton
number regardless of the accuracy of the solution to the parabolic
subsystem.

The outer iteration for solving equations (\ref{eq:drhoe-i}) --
(\ref{eq:dEg-i}) uses the following convergence criteria:
\begin{align}
  |T^{(k+1)}-T^{(k)}| < {} & \epsilon \, T^{(k+1)}, \label{eq:crT} \\
  |Y_e^{(k+1)}-Y_e^{(k)}| < {} & \epsilon \,
  Y_e^{(k+1)}, \label{eq:crY} \\
  |F_e^{(k+1)}| < {} & \epsilon \, \rho \left[|(\frac{\partial e}{\partial
      T})^{(k+1)}| |T^{(k+1)}-T^{(k)}| 
  + |(\frac{\partial e}{\partial
      Y_e})^{(k+1)}| |Y_e^{(k+1)}-Y_e^{(k)}| \right], \label{eq:crFe} \\
  |F_Y^{(k+1)}| < {} & \epsilon \, \rho Y_e^{(k+1)}, \label{eq:crFy}
\end{align}
where $\epsilon$ is a small number such as $\epsilon = 10^{-6}$.  We
do not use the relative error of the matter specific internal energy
$e$ because it may not accurately reflect the error since that $e$ can
be shifted by an arbitrary amount.  Typically, the outer iteration is
stopped only when all four conditions (Inequalities~\ref{eq:crT} --
\ref{eq:crFy}) are satisfied.  But for problems with extremely high
opacities, Inequalities~(\ref{eq:crFe}) \& (\ref{eq:crFy}) can be
difficult to achieve, and only Inequalities~(\ref{eq:crT})
\& (\ref{eq:crY}) are used.

\subsubsection{Acceleration Scheme}
\label{sec:accel}

We solve the multigroup diffusion equation (Equation~\ref{eq:MGdiff0})
using an iterative method.  The system of the multigroup radiation
diffusion equations is coupled.  Each individual radiation group
depends on the matter, which in turn depends on all radiation groups.
When the coupling is strong (e.g., large opacity and small heat
capacity) and/or the time step is large, the convergence rate of the
iterative method we have presented in \S~\ref{sec:innerIter} can be
very slow.  To improve the convergence rate, we have extended the
synthetic acceleration scheme of \citet{MorelLM85,MorelYW07} for photon
radiation to neutrino radiation, and have developed two types of
acceleration schemes.

\paragraph{Local Acceleration Scheme} 

We define the error of radiation energy density after $\ell$ inner
iterations as
\begin{equation}
  \epsilon_g^{\ell+1} = E_g^{e} - E_g^{\ell+1},
\end{equation}
where $E_g^e$ is the exact solution of Equation~(\ref{eq:MGdiff0}) and
$E_g^{\ell+1}$ is the result of the $\ell$-th inner iteration via
Equation~(\ref{eq:MGdiff1}).  In our local acceleration scheme, the
spatial derivatives of $\epsilon_g^{\ell+1}$ are assumed to be zero.
This is justified because the spatially constant limit corresponds to
the most slowly convergent mode \citep{MorelLM85} and our goal is to
improve the convergence rate by curing the slowest decaying mode of
the errors.  Using equations (\ref{eq:MGdiff0}) and
(\ref{eq:MGdiff1}), we derive an equation for the error in the $g$-th
group,
\begin{equation}
  \kappa_g^t \epsilon_g^{\ell+1} 
  -  H_g \sum_{g^{\prime}} \kappa_{g^{\prime}}
  \epsilon_{g^{\prime}}^{\ell+1} 
  -  \Theta_g \sum_{g^{\prime}} \xi_{g^\prime} \kappa_{g^{\prime}}
  \epsilon_{g^{\prime}}^{\ell+1} 
  =  H_g r_T^{\ell+1} + \Theta_g r_Y^{\ell+1}, \label{eq:acc1}
\end{equation}
where 
\begin{align}
  \kappa_g^t = { } & \kappa_g + \frac{1}{c\Delta t}, \\
  r_T^{\ell+1} = { } &\sum_{g^{\prime}} \kappa_{g^{\prime}}
  (E_{g^{\prime}}^{\ell+1} -E_{g^{\prime}}^{\ell}), \\
  r_Y^{\ell+1} = { } &\sum_{g^{\prime}} \xi_{g^\prime} \kappa_{g^{\prime}}
  (E_{g^{\prime}}^{\ell+1} -E_{g^{\prime}}^{\ell}).
\end{align}
Eq.~\ref{eq:acc1} can be solved analytically, and gives
\begin{equation}
  \epsilon_g^{\ell+1} = \frac{(qH_g + r \Theta_g) r_T^{\ell+1}+(p
    \Theta_g + sH_g) r_Y^{\ell+1}}{\kappa_g^t (pq-rs)}, \label{eq:local-a}
\end{equation}
where
\begin{align}
  p = { } & 1 - \sum_g \frac{H_g \kappa_g}{\kappa_g^t},\\
  q = { } & 1 - \sum_g \frac{\Theta_g \xi_g \kappa_g}{\kappa_g^t},\\
  r = { } & \sum \frac{H_g \xi_g \kappa_g}{\kappa_g^t},\\
  s = { } & \sum \frac{\Theta_g \kappa_g}{\kappa_g^t}.
\end{align}
The improved solution is therefore
\begin{equation}
  E_{g}^{\ell+1} := E_{g}^{\ell+1} + \epsilon_g^{\ell+1}.
\end{equation}
The acceleration is placed after the convergence check for the inner
iteration so that no unnecessary acceleration is applied to converged
solutions.

\paragraph{Gray Acceleration Scheme}

The original acceleration scheme of \citet{MorelLM85} utilizes a gray
diffusion equation for errors.  It is postulated that the normalized
multigroup spectrum of the errors is given by the normalized
equilibrium spectrum.  Note that our local acceleration scheme is
based on the assumption of local equilibrium.  Therefore the
spectrum defined as
\begin{equation}
  \psi_g = \frac{{\epsilon}_g^{\ell+1}}{\sum_{g^{\prime}}
    {\epsilon}_{g^{\prime}}^{\ell+1}},
\end{equation}
is postulated to be 
\begin{equation}
  \psi_g = \frac{\tilde{\epsilon}_g^{\ell+1}}{\sum_{g^{\prime}}
    \tilde{\epsilon}_{g^{\prime}}^{\ell+1}},
\end{equation}
where $\tilde{\epsilon}_g^{\ell+1}$ is the solution of the local
acceleration scheme (Equation~\ref{eq:local-a}).  Substituting
$E_g^{\ell+1} = E_g^e - \epsilon_g^{\ell+1}$ into Eq.~\ref{eq:MGdiff1}
and summing over groups, we obtain
\begin{equation}
  A \epsilon^{\ell+1} - \nabla \cdot (\bar{d} \nabla \epsilon^{\ell+1}) + \nabla
  \cdot (\vec{\phi} \epsilon^{\ell+1}) = c H r_T^{\ell+1} + c \Theta
  r_Y^{\ell+1}, \label{eq:gray-a}
\end{equation}
where
\begin{align}
  \epsilon^{\ell+1} = {} & {\sum_{g} {\epsilon}_{g}^{\ell+1}}, \\
  A = { } & c(1-H)\bar{\kappa} - c \Theta \bar{\kappa}_Y +
  \frac{1}{\Delta t},\\
  \bar{d} = { } & \sum_{g} \psi_g \frac{c\lambda_g}{\chi_g}, \\
  \vec{\phi} = { } & - \sum_{g} \frac{c\lambda_g}{\chi_g} \nabla \psi_g ,\\
  \bar{\kappa} = {} & \sum_g \psi_g \kappa_g, \\
  \bar{\kappa}_Y = {} & \sum_g \psi_g \xi_g \kappa_g.
\end{align}
Equation~(\ref{eq:gray-a}) is in the canonical form that CASTRO solves
(Equation~\ref{eq:canon}).  The solution of the gray diffusion
equation for error, $\epsilon^{\ell+1}$, is used to obtain the improved
multigroup solution
\begin{equation}
  E_{g}^{\ell+1} := E_{g}^{\ell+1} + \psi_g \epsilon^{\ell+1}.
\end{equation}
We note again that the acceleration operation is placed after the
convergence check for the inner iteration so that no unnecessary
acceleration is applied to converged solutions.

\section{Test Problems}
\label{sec:tests}

In this section we present detailed tests of the code demonstrating
its ability to handle a wide range of radiation hydrodynamics
problems. Photon radiation problems are presented in
Section~\ref{test-p}, whereas Section~\ref{test-n} shows neutrino
radiation hydrodynamics simulations of core-collapse supernovae.  Note
that not every term or equation in the full system of radiation
hydrodynamics is included in every test.  This allows various parts of
the code to be tested separately.

A CFL number of 0.8 is used for these tests unless stated otherwise or
a fixed time step is used.  For photon radiation test problems, no
additional constraint is posed by the implicit solver.  For
core-collapse supernova simulations, the time step is further
constrained by limiting the estimated relative changes in temperature
to less than 1\% and the estimated absolute changes in electron
fraction to less than 0.01 using the change rates from the previous
time step, and in our simulations this only happens around the time of
core bounce.  The relative tolerances for both the outer and the inner
iterations in the implicit update are $10^{-6}$.  All convergence
criteria for the outer iteration (Inequalities~\ref{eq:crT} --
\ref{eq:crFy}) are used except for the tests in
Sections~\ref{test:shocktube} and \ref{test:hse} with extremely high
opacities.  By default, the radiation groups are uniform in
logarithmic frequency space, and the representative frequency for the
$g$-th group is $\nu_g = \sqrt{\nu_{g-1/2}\nu_{g+1/2}}$.  For AMR
simulations, the ``deferred sync'' scheme (Paper II) is used for flux
synchronization between coarse and fine levels in the implicit solver.

\subsection{Photon Radiation Test Problems}
\label{test-p}

\subsubsection{Linear Multigroup Diffusion}
\label{test:lMGD}

In this problem, we simulate a linear multigroup diffusion test
problem introduced by \citet{ShestakovBolstab05}.  This can test the
multigroup diffusion part of the system
\begin{align}
\frac{\partial (\rho e)}{\partial t} = { } & \sum_{g} c
  (\kappa_gE_{g}-j_g) \label{eq:drhoe-mgd}\\
\frac{\partial E_g}{\partial t} = { } & -c (\kappa_gE_{g}-j_g)
  + \nabla \cdot \left(\frac{c\lambda_g}{\chi_g} \nabla E_g \right)
  . \label{eq:dEg-mgd}
\end{align}
To make the nonlinear system
mathematically tractable, \citet{ShestakovBolstab05} linearize the
system using the following assumptions.  First, the absorption
coefficient is assumed to be proportional to $\nu^{-3}$, and the
group-averaged absorption is given by
\begin{equation}
  \kappa_g = C_\kappa \nu_g^{-3},
\end{equation}
where $C_\kappa$ is a constant.  Second, radiation emission is assumed
to obey Wien's law rather than Planck's law, and it is further
modified so that
\begin{equation}
  B_g = \frac{2\pi h}{c^2} \nu_g^3 \left[\exp\left(-\frac{h\nu_{g-1/2}}{k T_{\mathrm{f}}}\right)
  - \exp\left(-\frac{h\nu_{g+1/2}}{k T_{\mathrm{f}}}\right)\right] T, \label{eq:Bg-LMD}
\end{equation}
where $T_{\mathrm{f}}$ is a fixed temperature.  Note that the
group-integrated $B_g$ now has linear dependence on $T$.  Finally,
there is no scattering, and the diffusion is not flux limited.  Under
these assumptions, \citet{ShestakovBolstab05} have obtained exact
solutions for the linearized system.

\begin{figure}
\epsscale{0.5}
\plotone{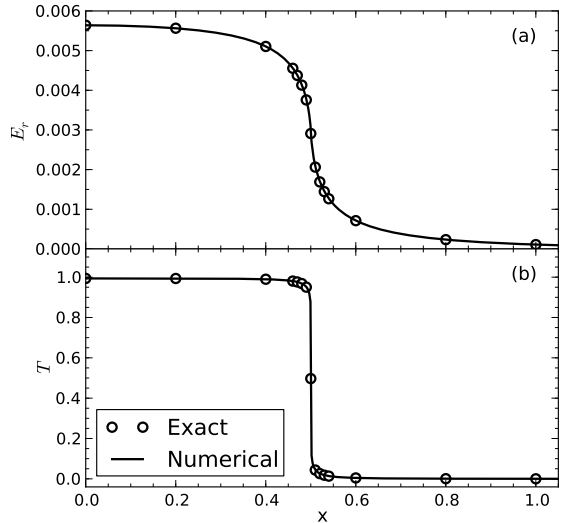}
\caption{ Linear multigroup diffusion test.  Numerical results are
  shown in lines, whereas the analytic solutions are shown in circle
  symbols.  We show dimensionless total radiation energy density
  and temperature at the dimensionless time of 1 in panels (a) and
  (b), respectively.  The spatial coordinate $x$ is also dimensionless. 
  \label{fig:SB}}
\end{figure}

\begin{deluxetable}{cccccc}
\tablecaption{Relative errors of linear multigroup diffusion test
problem.  The relative errors of temperature and
total radiation energy density at the dimensionless time of 1 are shown.
\label{tab:SB}}
\tablewidth{0pt}
\tablehead{
\colhead{$x$\tablenotemark{a}}    &   \colhead{$\epsilon(T) \times 10^3$}  &  
\colhead{$\epsilon(E_r) \times 10^3$} & 
\colhead{$x$}    &   \colhead{$\epsilon(T) \times 10^3$}  &  
\colhead{$\epsilon(E_r) \times 10^3$}
}
\startdata
0.00 & 0.0017 & 0.3209 & 0.51 & 4.8283 & 0.2585 \\
0.20 & 0.0016 & 0.3220 & 0.52 & 1.8030 & 0.0232 \\
0.40 & 0.0007 & 0.3460 & 0.53 & 1.0335 & 0.1495 \\
0.46 & 0.0077 & 0.4095 & 0.54 & 0.7124 & 0.2330 \\
0.47 & 0.0170 & 0.4443 & 0.60 & 0.3104 & 0.5469 \\
0.48 & 0.0461 & 0.5136 & 0.80 & 0.5774 & 1.4067 \\
0.49 & 0.2195 & 0.7169 & 1.00 & 1.3612 & 2.2413 \\
0.50 & 0.0020 & 0.3712 &      &        &
\enddata
\tablenotetext{a}{Dimensionless spatial coordinate}
\end{deluxetable}

The test we have simulated is the first problem in
\citet{ShestakovBolstab05}.  We have set up the test using physical
units rather than the dimensionless units of
\citet{ShestakovOffner08}.  The one-dimensional computational domain
in physical units is $(0, 1.0285926\times 10^{6})\,\mathrm{cm}$,
corresponding to $(0,5.12)$ in dimensionless units.  We use a
symmetry boundary condition at the lower boundary, and a Marshak
boundary with no incoming flux at the upper boundary (i.e., $E_g +
(2/3\kappa_g) \partial E_g/\partial x = 0$).  The mass density is
$1.8212111 \times 10^{-5}\,\mathrm{g}\,\mathrm{cm}^{-3}$.  The
specific heat capacity of the matter at constant volume is $9.9968637
\times 10^{7}\,\mathrm{erg}\,\mathrm{K}^{-1} \mathrm{g}^{-1}$.
Initially, the temperature is set to $0.1\,\mathrm{keV}$ and $0$, for
${x}<0.5$ and ${x}>0.5$, respectively.  Here, ${x}$ is a
dimensionless coordinate.  The radiation energy density is initially
set to zero everywhere.  The constant $C_\kappa$ is set to $4.0628337
\times 10^{43}\,\mathrm{cm}^{-1}\mathrm{Hz}^3$. The fixed temperature
in the group-integrated radiation emission (Equation~\ref{eq:Bg-LMD}) is
$T_{\mathrm{f}} = 0.01\,\mathrm{keV}$, corresponding to
$T_{\mathrm{f}} = 0.1$ in dimensionless units.  We use 64
radiation groups with the lowest boundary at zero.  The width of the
first group is set to $1.2089946 \times 10^{13}\,\mathrm{Hz}$,
corresponding to $5 \times 10^{-4}$ in dimensionless units, and
the width of other groups is set to be 1.1 times the width of its
immediately preceding group.  The representative frequency for each
group is set to $\nu_g = \sqrt{\nu_{g-1/2}\,\nu_{g+1/2}}$, except that
$\nu_1$ is set to $0.5\, \nu_{3/2}$ for the first group. We have run a
simulation with 2048 uniform cells.  Thus the dimensionless cell size
is $1/400$.  The time step is fixed at $5.8034112 \times
10^{-8}\,\mathrm{s}$, corresponding to $1/200$ in dimensionless
units.  No acceleration scheme is used in this test.
Figure~\ref{fig:SB} shows the results after 200 steps.  The comparison
with the tabular data of \citet{ShestakovOffner08} is shown in
Table~\ref{tab:SB}.  Relative errors are computed as $\epsilon(f) =
|f_{\mathrm{n}} - f_{\mathrm{e}}|/ f_{\mathrm{e}}$, where
$f_{\mathrm{n}}$ and $f_{\mathrm{e}}$ are the numerical and exact
results, respectively.  Because the tabular data of the exact solution
are located on the faces of numerical cells, averaging of the
numerical data has been performed for the comparison.  The results are
in excellent agreement with the exact solution, and are comparable to
the numerical results of \citet{ShestakovOffner08}.

\begin{deluxetable}{cccc}
  \tablecaption{$L_1$-norm errors and convergence rate for the total
    radiation energy density at the dimensionless time of 1 in the
    linear multigroup diffusion test problem.  Four one-dimensional
    uniform-grid runs with various resolutions are shown.  The errors
    are computed on the dimensionless domain of $(0,1)$.
\label{tab:uniform}}
\tablewidth{0pt}
\tablehead{
\colhead{$\Delta x$} & \colhead{$\Delta t$}    &   \colhead{$L_1$ Error}  &  
\colhead{Convergence Rate} 
}
\startdata
1/100 &  2/25  & 2.91E-3 &     \\
1/200 &  1/50  & 7.36E-4 & 2.0 \\
1/400 &  1/200 & 1.90E-4 & 1.9 \\
1/800 &  1/800 & 4.87E-5 & 2.0 
\enddata
\end{deluxetable}

\begin{deluxetable}{cccc}
  \tablecaption{$L_1$-norm errors and convergence rate for the total
    radiation energy density at the dimensionless time of 1 in the
    linear multigroup diffusion test problem.  Four one-dimensional
    AMR runs with various resolutions are shown.  The errors
    are computed on the dimensionless domain of $(0,1)$.
\label{tab:amr}}
\tablewidth{0pt}
\tablehead{
\colhead{$\Delta x$\tablenotemark{a}} & \colhead{$\Delta t$\tablenotemark{b}}
&   \colhead{$L_1$ Error}  &  
\colhead{Convergence Rate} 
}
\startdata
1/50, 1/100 & 4/25, 2/25  & 3.58E-3 &    \\
1/100, 1/200 & 1/25, 1/50  & 1.05E-3 & 1.8 \\
1/200, 1/400 & 1/100, 1/200 & 2.69E-4 & 2.0 \\
1/400, 1/800 & 1/400, 1/800 & 6.82E-5 & 2.0
\enddata
\tablenotetext{a}{Cell size on levels 0 and 1}
\tablenotetext{b}{Time step on levels 0 and 1}
\end{deluxetable}

To assess the convergence behavior of our implicit scheme, we have
performed a series of uniform-grid and AMR simulations of the linear
multigroup diffusion test problem with various resolutions.  The
initial setup is the same as above except for spatial resolution and
time step.  Two levels are used in the AMR runs with the dimensionless
domain of (0.25,0.75) covered by the fine (i.e., level 1) grids.  In
the convergence study, when the spatial resolution changes by a factor
of 2 from one run to another, we change the time step by a factor of
4.  Because our implicit scheme is first-order in time and
second-order in space, the expected convergence rate in the
convergence study is second-order with respect to $\Delta x$.
Tables~(\ref{tab:uniform}) and (\ref{tab:amr}) show the $L_1$-norm
errors and convergence rate for the total radiation energy density on
the dimensionless spatial domain of $(0,1)$ at the dimensionless time
of 1.  Here, the $L_1$-norm error of the total radiation energy
density is computed as
\begin{equation}
  \epsilon_1(E_r) = \frac{ \sum_{i} |E_{r,i}-E_r(x_i)|
\Delta x_{i} }{ \sum_{i} E_r(x_i) \Delta x_{i}},
\end{equation}
where $E_{r,i} = \sum_g E_{g,i}$ is the numerical result for cell $i$
and $E_r(x_i)$ is the ``exact'' solution obtained from a
high-resolution uniform-grid simulation with $\Delta x = 1/3200$ and
$\Delta t = 1/12800$.  The results of both uniform-grid and AMR
simulations demonstrate second-order convergence rate with respect to
$\Delta x$ as expected.  Because only part of the domain in the AMR
runs is covered by fine grids, the AMR runs have somewhat larger
errors than the uniform-grid runs with the same resolution as the fine
grids.

\subsubsection{Nonequilibrium Radiative Transfer With Picket-Fence Model}

\citet{SuOlson99} have studied a nonequilibrium radiative transfer
problem with a multigroup model, more specifically the picket-fence
model.  There are some special assumptions in this test problem.  The
volumetric heat capacity at constant volume is assumed to be $c_v =
\alpha T^3$, where $T$ is the matter temperature and $\alpha$ is a
parameter.  The group-integrated Planck function is assumed to be
\begin{equation}
  B_g = p_g \left(\frac{a c}{4 \pi}\right) T^4,
\end{equation}
where $a$ is the radiation constant, and $p_g$ are parameters such
that $\sum_{g}p_g = 1$.  It is also assumed that the absorption
coefficient is independent of temperature and there is no scattering.
\citet{SuOlson99} have also defined dimensionless variables for
convenience.  The dimensionless spatial coordinate is defined as
\begin{equation}
  x = \bar{\kappa} z,
\end{equation}
where $z$ is the coordinate in physical units, and 
\begin{equation}
  \bar{\kappa} = \sum_{g}p_g \kappa_g.
\end{equation}
The dimensionless time is defined as
\begin{equation}
  \tau = \left(\frac{4 a c \bar{\kappa}}{\alpha}\right) t,
\end{equation}
where $t$ is the time in physical units.  The dimensionless variables
$U_g$ for radiation energy density and $V$ for matter energy density
are defined as
\begin{eqnarray}
  U_g &=& \frac{E_g}{a T_0^4}, \\
  V &=& \left(\frac{T}{T_0}\right)^4,
\end{eqnarray}
where $T_0$ is a reference temperature.  In this test problem, there
is a constant radiation source that exists for a finite period of time
$(0\le\tau \le \tau_0)$ on a finite range $(|x|<x_0)$.
\citet{SuOlson99} have found analytic solutions to the problem for
both transport and diffusion approaches.

\begin{figure}
\epsscale{0.5}
\plotone{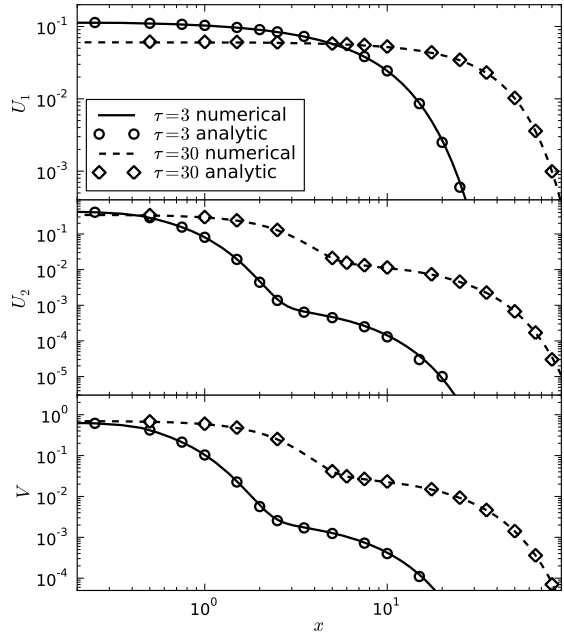}
\caption{ Nonequilibrium radiative transfer with picket-fence model.
  Numerical results of $U_1$, $U_2$, and $V$ are shown for $\tau = 3$
  ({\it solid lines}) and $30$ ({\it dashed lines}).  Also shown are
  analytic results for $\tau = 3$ ({\it circles}) and $30$ ({\it
    diamonds}).
  \label{fig:SO}}
\end{figure}

Here, we simulate the Case C of \citet{SuOlson99} by solving the
multigroup diffusion equations (Equations~\ref{eq:drhoe-mgd} \&
\ref{eq:dEg-mgd}), and compare the results with the diffusion solution
of \citet{SuOlson99}.  Two radiation groups are used in this test,
with $p_1 = p_2 = 0.5$.  The absorption coefficients are set to
$\kappa_1 = 2/101\,\mathrm{cm}^{-1}$ and $\kappa_2 =
200/101\,\mathrm{cm}^{-1}$.  Thus, $\bar{\kappa} =
1\,\mathrm{cm}^{-1}$.  The constant $\alpha$ in the expression for
volumetric heat capacity is set to $\alpha = 4 a$.  The reference
temperature is chosen as $T_0 = 10^6\,\mathrm{K}$.  The parameters for
the radiation source are $x_0 = 0.5$, and $\tau_0 = 10$.  When $\tau <
\tau_0$, the radiation source deposits radiation energy to the region
$|x| < x_0$ and increases the radiation energy density at a rate of
$p_gc\bar{\kappa}aT_0^4$ for each group.  The computational domain is
$0<x<102.4$ covered by 1024 uniform cells.  The lower boundary is
symmetric, and the radiation energy density outside the upper boundary
is set to zero.  Initially, there is no radiation and the temperature
is zero.  A fixed time step $\Delta \tau = 0.1$ is employed.  The
explicit solver is turned off in the simulation, and the radiation
flux is not limited. No acceleration scheme is used in this test.  The
numerical results are shown in Figure~\ref{fig:SO}.  Our results are
in excellent agreement with the analytic solution.

\subsubsection{Radiating Sphere}
\label{sec:RadSphere}

\citet{Graziani08} found an analytic solution for the spectrum of
a radiating sphere in a medium with frequency-dependent opacity.  This
test was used by \citet{SwestyMyra09} to test their V2D code. In
this test, a hot sphere with a fixed temperature is surrounded by a
cold static medium.  The absorption coefficient is assumed to be zero,
and the radiative transfer is due to coherent and isotropic
scattering.  Thus, there is no radiation-matter coupling or group
coupling.  The time-dependent spectrum at a given place in the medium
is therefore the superposition of the black-body spectrum of the
medium and the spectrum of the radiation originated from the hot
sphere.

\begin{figure}
\epsscale{0.5}
\plotone{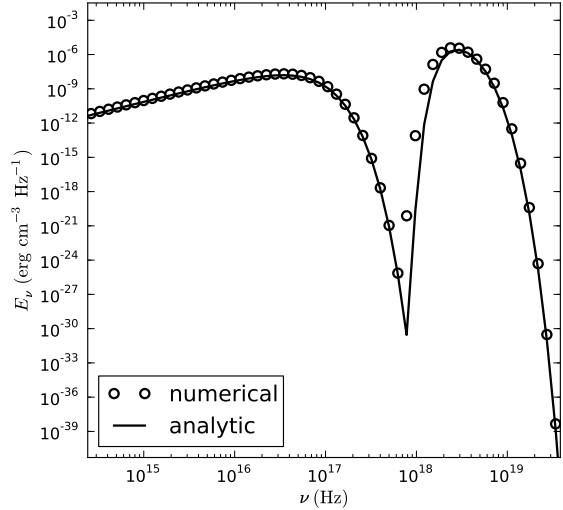}
\caption{ Spectrum from the radiating sphere observed at $t =
  10^{-12}\,\mathrm{cm}$ and $r = 0.06\,\mathrm{cm}$.  Both numerical
  ({\it circle symbols}) and analytic ({\it
    solid line}) results are shown.
  \label{fig:radsphere}}
\end{figure}

We perform this test in one-dimensional spherical geometry.  The
computational domain is $0.02\,\mathrm{cm} < r < 0.1\,\mathrm{cm}$ and
is covered by 128 uniform cells.  The hot sphere has a fixed
temperature of $1.5\,\mathrm{keV}$, and the medium is at $50\,\mathrm{eV}$.
We use 60 energy groups covering an energy range of $0.5\,\mathrm{eV}$
to $308\,\mathrm{keV}$.  The group Rosseland mean coefficient is
assumed to be $\chi_g = 10^{13} (\nu_g/3.6\times
10^{14}\,\mathrm{Hz})^{-3}\,\mathrm{cm}^{-1}$.  The explicit solver is
turned off.  No acceleration scheme is used in this test.  There is no
outer iteration in the implicit solver because the matter properties
do not change.  The time step is fixed at $10^{-15}\,\mathrm{s}$.  The
spectrum at $r = 0.06\,\mathrm{cm}$ and $t = 10^{-12}\,\mathrm{s}$ is
shown in Figure~\ref{fig:radsphere}.  The numerical results are again in
good agreement with the analytic solution.

\subsubsection{Shock Tube Problem In Strong Equilibrium Regime}
\label{test:shocktube}

\begin{figure}
\epsscale{0.6}
\plotone{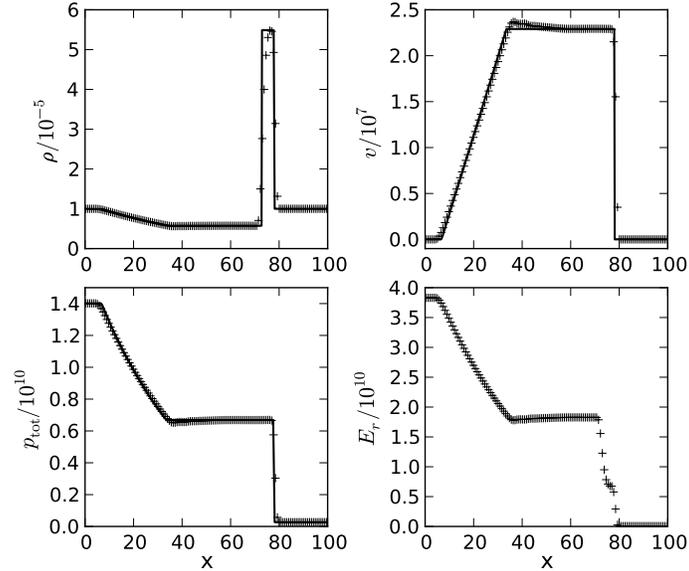}
\caption{ Shock tube at $t = 10^{-6}\,\mathrm{s}$.  Numerical results
  from a full radiation hydrodynamics calculation with 16 radiation
  groups and  128 cells are
  shown in symbols.  The exact solutions of a
  pure hydrodynamic Riemann problem corresponding to the numerical
  problem are shown in solid lines for comparison.  We show mass
  density ($\rho$), velocity 
  ($v$), total pressure ($p_{\mathrm{tot}}$), and radiation energy
  density ($E_r$).  For the radiation hydrodynamics simulation, the
  total pressure is the sum of gas pressure and 
  radiation pressure, and the radiation energy density is summed
  over all groups.  In the pure hydrodynamic Riemann problem, there is
  no radiation energy density.  All quantities are in cgs units.
  \label{fig:shocktube}}
\end{figure}

In this one-dimensional problem, the initial conditions of the matter
are $\rho_L = 10^{-5}\,\mathrm{g}\,\mathrm{cm}^{-3}$, $T_L = 1.5
\times 10^{6}\,\mathrm{K}$, $u_L = 0$ and $\rho_R =
10^{-5}\,\mathrm{g}\,\mathrm{cm}^{-3}$, $T_R = 3 \times
10^{5}\,\mathrm{K}$, $u_R = 0$, where $L$ stands for the left half,
and $R$ the right half of the computational domain ($0 < x <
100\,\mathrm{cm}$).  The gas is assumed to be ideal with an adiabatic
index of $\gamma = 4/3$ and a mean molecular weight of $\mu = 1$.
Initially, the radiation is assumed to be in thermal equilibrium with
the gas (i.e., $E_g = 4 \pi B_g / c$).  The interaction coefficients
are set to $\kappa_{g} = 10^6\,\mathrm{cm}^{-1}$ and
$\chi_{\mathrm{g}} = 10^{8}\,\mathrm{cm}^{-1}$. Thus, due to the huge
opacities, the system is close to the limit of strong equilibrium with
no diffusion, and is essentially governed by Equations
(\ref{eq:hyper-rho})--(\ref{eq:hyper-Eg}) with $\lambda_g = 1/3$ and
$f_g = 1/3$.  The parameters of this test problem are chosen such that
neither radiation pressure nor gas pressure can be ignored.  This test
is adapted from the shock tube problem in Paper II.  However, the
adiabatic index of the gas is now $\gamma = 4/3$ rather than $5/3$ so
that we can compare the numerical results with the exact solution of a
pure hydrodynamic Riemann problem.  Furthermore, although this is a
one-temperature gray problem due to the huge opacity, the numerical
calculation is performed with 16 radiation groups with the lowest and
uppermost boundaries at $10^{14}\,\mathrm{Hz}$ and
$10^{18}\,\mathrm{Hz}$, respectively.  We use 128 uniform zones in
this test.  The numerical results are shown in
Figure~\ref{fig:shocktube}.  Also shown are the exact solutions of a
pure hydrodynamic Riemann problem corresponding to the numerical
problem.  The results show good agreement.  We also note that our
scheme is stable without using small time steps even though the system
is in the ``dynamic diffusion'' limit \citep{MihalasMihalas99} with
$\tau u/c \sim 10^{7}$, where $\tau$ is the optical depth of the
system.  Either the gray acceleration scheme or the local acceleration
scheme (\S~\ref{sec:accel}) must be employed in this test, otherwise
the implicit solver, which is responsible for the thermal equilibrium
between radiation and matter, exhibits extremely slow convergence. 

\subsubsection{Nonequilibrium Radiative Shock}
\label{test:2Tshock}

\begin{figure}
\epsscale{0.5}
\plotone{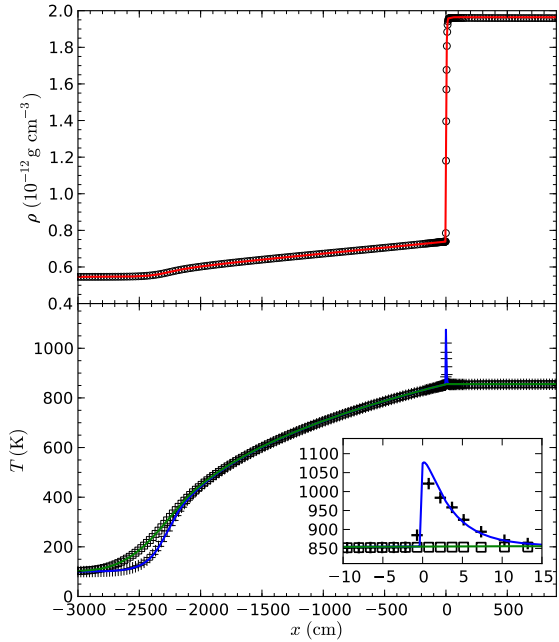}
\caption{ Density and temperature profiles for Mach 5 supercritical
  shock.  Numerical results are shown in symbols.  We show density
  ({\it circles}), gas temperature ({\it plus signs}), and radiation
  temperature ({\it squares}).  Here radiation temperature is defined
  as $(E_r/a)^{1/4}$, where $E_r$ is the total radiation energy
  density.  Also shown are the analytic results for density 
  ({\it red solid line}), gas temperature ({\it blue solid line}), and
  radiation temperature ({\it green solid line}). We show only part of
  the region near the density jump.  The inset shows a blow-up of the
  spike in temperature.  The
  numerical results have been shifted by $-207\,\mathrm{cm}$
  in space to compensate for the discrepancy in shock position caused
  by the initial setup.
  \label{fig:M5}}
\end{figure}

In this test, we simulate the Mach 5 shock problem of
\citet{LowrieEdwards08}.  In Paper II, we showed results
computed with the gray radiation hydrodynamics solver.  Here, we
present the numerical results obtained with the multigroup radiation
hydrodynamics solver.  In this test, the one-dimensional numerical
region $(-4000\,\mathrm{cm}, 2000\,\mathrm{cm})$ initially consists of
two constant states: $\rho_L = 5.45969 \times
10^{-13}\,\mathrm{g}\,\mathrm{cm}^{-3}$, $T_L = 100\,\mathrm{K}$, $u_L
= 5.88588 \times 10^{5}\,\mathrm{cm}\,\mathrm{s}^{-1}$ and $\rho_R =
1.96435 \times 10^{-12}\,\mathrm{g}\,\mathrm{cm}^{-3}$, $T_R =
855.720\,\mathrm{K}$, $u_R = 1.63592 \times
10^{5}\,\mathrm{cm}\,\mathrm{s}^{-1}$, where $L$ stands for the left,
and $R$ the right of the point $x=0$, respectively.  The gas is
assumed to be ideal with an adiabatic index of $\gamma = 5/3$ and a
mean molecular weight of $\mu = 1$.  The setup is the same as that in
Paper II \footnote{Unlike the simulation shown in Paper II, the
  simulation here used the 2010 CODATA recommended values of the
  fundamental physical constants.  Therefore, the physical values of
  the states have changed slightly.} except that 16 radiation groups
are used in this paper.  The lowest and uppermost boundaries of the
groups are at $10^{10}\,\mathrm{Hz}$, and $10^{15}\,\mathrm{Hz}$,
respectively.  Initially, the radiation is assumed to be in thermal
equilibrium with the gas (i.e., $E_g = 4 \pi B_g / c$).  The
interaction coefficients are set to $\kappa_{g} = 3.92664 \times
10^{-5}\,\mathrm{cm}^{-1}$ and $\chi_{g} =
0.848903\,\mathrm{cm}^{-1}$, respectively.  The states at spatial
boundaries are held at fixed values.  Four refinement levels (five
total levels) are used with a refinement factor of 2 with the finest
resolution being $\Delta x \approx 1.5\,\mathrm{cm}$.  In this test, a
cell is tagged for refinement if the second derivative of either
density or temperature exceeds certain thresholds (Paper II).  The
initial conditions are set according to the pre-shock and post-shock
states of the Mach 5 shock.  After a brief period of adjustment, the
system settles down to a steady shock.  The simulation is stopped at
$t = 0.04\,\mathrm{s}$.  The results are shown in Figure~\ref{fig:M5}.
The agreement with the analytic solution including the narrow spike in
temperature is excellent, except that the numerical results in the
figure had to be shifted by $-207\,\mathrm{cm}$ in space to match the
analytic shock position.  This discrepancy in shock position is due to
the initial transient phase as the initial state relaxes to the
correct steady-state profile.  No flux limiter is used in this
calculation because the analytic solution of \citet{LowrieEdwards08}
assumes $\lambda = 1/3$.  We again found that the implicit solver would
fail to converge without an acceleration scheme.  Both the gray
acceleration scheme and the local acceleration scheme
(\S~\ref{sec:accel}) work in this test.

\subsubsection{Advecting Radiation Pulse}
\label{test:pulse}

In Paper II, we performed a test of advecting a radiation pulse by the
motion of matter with the gray radiation hydrodynamics solver
\citep[see also][]{KrumholzKMB07}.  Here, we repeat the test with
small changes to exercise the multigroup aspect of CASTRO.  The setup
is the same as that in Paper II, except that the absorption
coefficient $\kappa$ in this paper is given by
\begin{equation}
  \kappa = 180 \left(\frac{T}{10^7\,\mathrm{K}}\right)^{-0.5}
  \left(\frac{\nu}{10^{18}\,\mathrm{Hz}}\right)^{-3} \left[1 -
    \exp\left(-\frac{h\nu}{kT}\right)\right] \,\mathrm{cm}^{-1}. \label{eq:kappulse}
\end{equation}
As in the test in Paper II, the initial temperature and density
profiles are
\begin{eqnarray}
  T &=& T_0 + (T_1-T_0) \exp{\left(-\frac{x^2}{2w^2}\right)}, \label{eq:pulse}\\
  \rho &=& \rho_0\frac{T_0}{T_1} + \frac{a \mu}{3R} \left(\frac{T_0^4}{T} - T^3\right),
\end{eqnarray}
where $T_0 = 10^7\,\mathrm{K}$, $T_1=2\times 10^7\,\mathrm{K}$,
$\rho_0=1.2\,\mathrm{g\:cm}^{-3}$, $w = 24\,\mathrm{cm}$, the mean
molecular weight of the gas is $\mu=2.33$, and $R$ is the ideal gas
constant.  Initially the radiation is assumed to be in thermal
equilibrium with the gas.  No flux limiter is used in this test (i.e.,
$\lambda = 1/3$ and $f = 1/3$).  If there were no radiation diffusion,
the system would be in an equilibrium with the gas pressure and
radiation pressure balancing each other.  Because of radiation
diffusion, the balance is lost and the gas moves.

The purpose of this test is twofold.  First, we use this test to
assess the ability of the code to handle radiation being advected by
the motion of matter.  We solve the problem numerically in different
laboratory frames.  Then we compare the case in which the gas is
initially at rest to the case in which the gas initially moves at a
constant velocity.  Another purpose is to assess the effect of group
resolution.  Since the opacity is very large, the system is close to
thermal equilibrium between radiation and matter.  Hence, the results
of multigroup radiation hydrodynamics simulations should become
increasingly close to those of a gray radiation hydrodynamics
simulation as more groups are used.

\begin{figure}
\epsscale{0.5}
\plotone{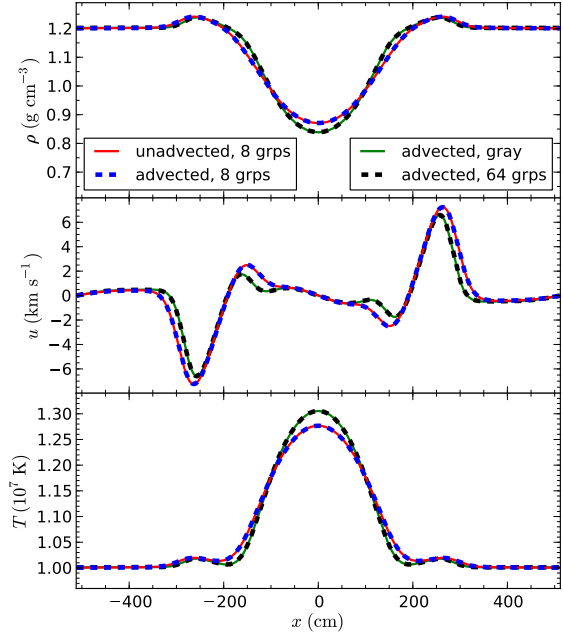}
\caption{ Density, velocity and temperature profiles at $t = 4.8
  \times 10^{-5}\,\mathrm{s}$ for the test of advecting radiation
  pulse.  We show the results of four runs including the unadvected
  8-group run ({\it red solid lines}), the advected 8-group run ({\it
    blue dotted lines}), the advected 64-group run ({\it black dotted
    lines}), and the advected gray run ({\it green solid lines}).  The
  results of the advected runs have been shifted in space for
  comparison. 
  \label{fig:pulse}}
\end{figure}

\begin{figure}
\epsscale{0.5}
\plotone{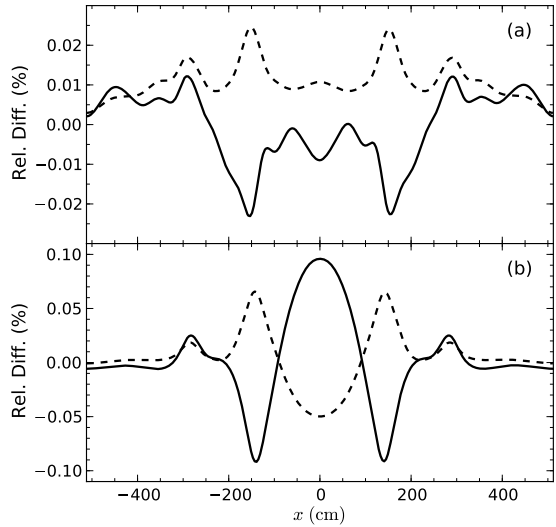}
\caption{ Relative differences in density ({\it solid lines}) and gas
  temperature ({\it dashed lines}) in the test of advecting radiation
  pulse.  The relative differences between the advected 8-group and
  unadvected 8-group runs are shown in panel (a), whereas the
  relative differences between the advected 64-group and advected gray runs in
  panel (b). 
  \label{fig:pulse-error}}
\end{figure}

We have performed four runs: an unadvected run with 8 radiation
groups, an advected run with 8 groups, an advected run with 64 groups,
and an advected run using the gray radiation hydrodynamics solver
presented in Paper II.  The velocity in the unadvected run is
initially zero everywhere, whereas in the advected runs it is
$10^6\,\mathrm{cm\:s}^{-1}$ everywhere.  For the multigroup runs, the
lowest and uppermost group boundaries are at $10^{15}\,\mathrm{Hz}$
and $10^{19}\,\mathrm{Hz}$, respectively.  In the gray simulation, the
single group Planck mean coefficient, $\kappa_{\mathrm{P}}$, and
Rosseland mean coefficient, $\kappa_{\mathrm{R}}$, of the opacity
given by Equation~(\ref{eq:kappulse}) are
\begin{eqnarray}
  \kappa_{\mathrm{P}} &=& 3063.96
  \left(\frac{T}{10^7\,\mathrm{K}}\right)^{-3.5}
  \,\mathrm{cm}^{-1}, \\
  \kappa_{\mathrm{R}} &=& 101.248
  \left(\frac{T}{10^7\,\mathrm{K}}\right)^{-3.5}
  \,\mathrm{cm}^{-1} .
\end{eqnarray}
The computational domain in all four runs is a one-dimensional region
of $-512\,\mathrm{cm} < x < 512\,\mathrm{cm}$ with periodic
boundaries.  The numerical grid consists of 512 uniform cells.  The
simulations are stopped at $t = 4.8 \times 10^{-5}\,\mathrm{s}$.
Figure~\ref{fig:pulse} shows the density, velocity, and temperature
profiles for all four runs.  The results of the advected runs have
been shifted in space for comparison.  The profiles from the
unadvected 8-group run and the advected 8-group run are almost
identical demonstrating the accuracy of our scheme in handling
radiation advection by the motion of matter. This is expected because
the velocity is significantly smaller than the speed of light ($v/c \sim
3 \times 10^{-5}$).  The results from the advected 64-group run and
the gray run are also almost identical as expected.  It is clear
that the 8-group results differ from the gray results.  This is not
surprising because we did not perform group averaging for interaction
coefficients (Equations~\ref{eq:kappa-g} \& \ref{eq:chi-g}).  We also
show the relative differences in Figure~\ref{fig:pulse-error}.  The
relative difference of the advected 8-group run with respect to the
unadvected 8-group run is computed as (advected - unadvected) /
unadvected, and the relative difference of the 64-group run with
respect to the gray run is computed as (multigroup - gray) / gray.
The figure shows that the difference between the advected 8-group and
unadvected 8-group runs is less than 0.03\% everywhere, and the
difference between the 64-group and gray runs is less than 0.1\%
everywhere.

\subsubsection{Static Equilibrium}
\label{test:hse}

\begin{figure}
\epsscale{0.6}
\plotone{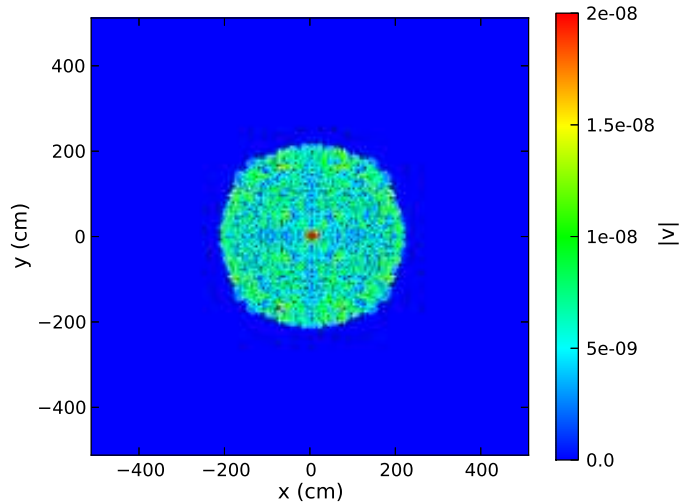}
\caption{ Structure of the magnitude of velocity at $t = 10^{-4}
  \,\mathrm{s}$ in the static equilibrium test.  The unit of velocity
  is $\mathrm{cm\:s}^{-1}$.
  \label{fig:pulse-hse}}
\end{figure}

In Paper II, we have demonstrated that the gray solver can
maintain a perfect static equilibrium in multiple dimensions because the
radiation pressure and the gas pressure are coupled in the Riemann
solver.  Here, we repeat the test using the multigroup solver.  The
setup is the same as that in Paper II.  The initial conditions for
radiation and matter are the same as those in the test of the advecting
radiation pulse (\S~\ref{test:pulse}) except that the interaction
coefficients are set to a very high value, $\kappa_g = \chi_g =
10^{20}\,\mathrm{cm}^2\,\mathrm{g}^{-1} \times \rho$.  Thus, almost no
diffusion can happen, and the radiation and the matter are in perfect
thermal equilibrium.  We have performed a calculation on a
two-dimensional Cartesian grid of $-512\,\mathrm{cm} < x <
512\,\mathrm{cm}$ and $-512\,\mathrm{cm} < y < 512\,\mathrm{cm}$ with
512 uniform cells in each direction.  The initial velocity is zero
everywhere.  For the initial setup, the coordinate $x$ in equation
(\ref{eq:pulse}) is replaced by $r = \sqrt{x^2 + y^2}$.  We use 16
radiation groups with the lowest and uppermost boundaries at
$10^{15}\,\mathrm{Hz}$ and $10^{19}\,\mathrm{Hz}$, respectively.
Figure~\ref{fig:pulse-hse} shows the velocity profile at $t = 10^{-4}
\,\mathrm{s}$ (after about 6000 steps).  The maximal velocity at that
time is $\sim 8 \times 10^{-8}\,\mathrm{cm\:s}^{-1}$, which is about
$10^{-15}$ of the sound speed.  Such a small gas velocity indicates
that the multigroup solver in CASTRO can also maintain a perfect
static equilibrium in multiple dimensions, as does the gray solver in
CASTRO.

\subsection{Core-Collapse Supernova Simulations}
\label{test-n}

In this section, we present core-collapse supernova simulations in 1D
spherical and 2D cylindrical coordinates.  Newtonian gravity with a
monopole approximation is used in the 2D simulation for simplicity,
although CASTRO has a multigrid Poisson solver for gravity.  In the
simulations, there are three neutrino species: electron neutrinos,
$\nu_e$, electron antineutrinos, $\bar{\nu}_e$, and $\nu_\mu$ which
stands for a combined species representing $\nu_\mu$, $\nu_\tau$,
$\bar{\nu}_\mu$, and $\bar{\nu}_\tau$.  We use a tabular equation of
state that provides thermodynamic quantities as a function of
temperature, density, and electron fraction and was constructed from
the \citet{SHenEOSa,ShenEOSb} mean-field table, augmented with photons
and a general electron/positron equation of state, and extended down
to $10\:\mathrm{g}\:\mathrm{cm}^{-3}$.  The opacities used are fully
described in \citet{BurrowsRT06}.  They include 1) elastic scattering
off nucleons, alpha particles, and the single representative heavy
nucleus at the peak of the binding energy curve calculated by
\citet{SHenEOSa,ShenEOSb}, and 2) charged-current absorption cross
sections off nucleons and nuclei (the latter using the approach of
\citet{Bruenn85}). The scattering and absorption cross sections off
nucleons are corrected for weak magnetism and recoil effects, the form
factor, electron screening, and ion-ion correlation effects on
Freedman scattering off nuclei are applied, and pair production by
nucleon-nucleon bremsstrahlung and electron-positron annihilation (and
their inverses) are included as sources (and sinks).  The sink terms
of the latter do not include Fermi blocking corrections by the
neutrinos in the final state.  Inelastic neutrino-electron scattering
is treated as elastic scattering.

\subsubsection{One-Dimensional Core-Collapse Supernova Simulation}
\label{sec:ccsn1d}

We present here a core-collapse supernova simulation in
one-dimensional spherical coordinates.  The initial model is based on
a $15\,M_\odot$ progenitor of \citet{WoosleyWeaver95}.  The
computational domain for this run is $0<r<5\times 10^8\,\mathrm{cm}$.
The simulation employs four refinement levels (five total levels) with
1024 zones on the coarsest level (i.e., level 0) and a resolution of
$\Delta r \approx 0.305\,\mathrm{km}$ on the finest level (i.e., level
4).  The AMR refinement factor is 2.  The regions where the density is
greater than $7.2 \times 10^{6}\,\mathrm{g}\,\mathrm{cm}^{-3}$, or the
enclosed mass is greater than $1.32\,M_{\odot}$ are refined to level
2.  In addition, cells on level $\ell$ will be refined to level
$\ell+1$ if $Y_e < 0.4$, $s > 5$, $\nabla Y_e > 0.01/ \Delta r_\ell$,
or $\nabla s > 1 / \Delta r_\ell$, where $\ell < 4$, $s$ is entropy in
units of $k\,\mathrm{baryon}^{-1}$, and $\Delta r_\ell$ is the cell
width on level $\ell$.  We use 40 energy groups for electron neutrinos
with the first group at $1\,\mathrm{MeV}$ and the last group at
$300\,\mathrm{MeV}$.  We use 32 energy groups for electron
antineutrinos with the first and last groups at $1\,\mathrm{MeV}$ and
$100\,\mathrm{MeV}$, respectively.  For $\nu_\mu$, we use 40 energy
groups with the first and last groups at $1\,\mathrm{MeV}$ and
$300\,\mathrm{MeV}$, respectively.  The flux limiter of
\citet{LevermorePomraning81} is used is this simulation.  The local
acceleration scheme is employed.  The inner boundary is symmetric.  An
outflow boundary condition is used for the outer boundary in the
explicit solver.  The modified Marshak boundary of
\citet{SanchezPomraning91} with zero incoming flux is used for the
outer boundary in the implicit solver.  Thus, the region outside the
outer boundary is essentially treated as a vacuum, and the radiation
diffuses into the vacuum.  Note that in the usual Marshak boundary
based on the classical diffusion theory, the flux at the boundary does
not have the correct behavior in the streaming limit, and the issue
has been fixed in the modified Marshak boundary of
\citet{SanchezPomraning91} by using flux limited diffusion.

\begin{figure}
\epsscale{0.7}
\plotone{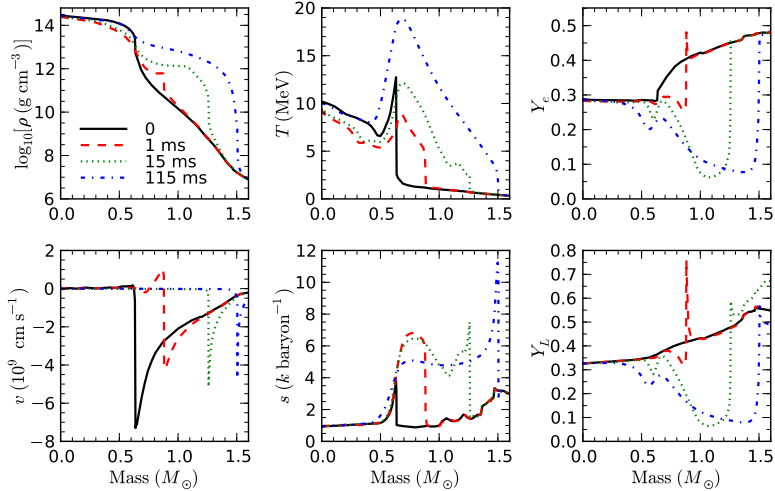}
\caption{ Density ({\it upper left}), velocity ({\it lower left}),
  temperature ({\it upper center}), entropy per baryon ({\it lower
    center}), electron fraction ({\it upper right}), and net lepton
  fraction ({\it lower right}) as a function of enclosed mass for the
  1D core-collapse supernova simulation.  Snapshots are shown for 0
  ({\it black solid lines}), 1 ms ({\it red dashed lines}), 15 ms
  ({\it green dotted lines}), and 115 ms ({\it blue dash-dotted
    lines}) after bounce. 
  \label{fig:ss1d}}
\end{figure}

The simulation is carried out to 200 ms after core bounce.
Figure~\ref{fig:ss1d} shows four snapshots of density, velocity,
temperature, entropy, electron fraction, and net electron lepton
fraction ($Y_L$).  The infall of the material due to gravity results
in the increase of the central density until the core bounces.  At
bounce, the density reaches $\rho^{\mathrm{core}} \approx 2.9 \times
10^{14}\:\mathrm{g}\:\mathrm{cm}^{-3}$ at the core.  A shock appears
around bounce at $M_{\mathrm{sh}} \approx 0.635 M_\odot$.  The
electron and lepton fractions at bounce are $Y_e^{\mathrm{core}}
\approx 0.288$ and $Y_L^{\mathrm{core}} \approx 0.326$, respectively,
at the core.  The core temperature at bounce is $T^{\mathrm{core}}
\approx 9.98\:\mathrm{MeV}$.  The core entropy at bounce is
$s^{\mathrm{core}} \approx 0.901\:k\:\mathrm{baryon}^{-1}$.  These
properties are in agreement with previous studies
\citep[e.g.,][]{RamppJanka00, Liebendorfer_etal01, ThompsonBP03}.  In
a recent study, \citet{Lentz_etal12} have obtained $M_{\mathrm{sh}}
\approx 0.492\:\mathrm{M_\odot}$, $\rho_{\mathrm{core}} \approx 4.264
\times 10^{14}\:\mathrm{g}\:\mathrm{cm}^{-3}$, $Y_e^{\mathrm{core}}
\approx 0.2407$, and $Y_L^{\mathrm{core}} \approx 0.2782$ for their
model N-FullOp at bounce, and $M_{\mathrm{sh}} \approx
0.717\:\mathrm{M_\odot}$, $\rho_{\mathrm{core}} \approx 3.336 \times
10^{14}\:\mathrm{g}\:\mathrm{cm}^{-3}$, $Y_e^{\mathrm{core}} \approx
0.3046$, and $Y_L^{\mathrm{core}} \approx 0.3696$ for their model
N-ReduceOp at bounce.  Except for $\rho_{\mathrm{core}}$, the
properties of our model at bounce are in between those in models
N-FullOp and N-ReduceOp of \citet{Lentz_etal12}.

Figure~\ref{fig:ss1d} shows a spike near the shock in the $Y_e$ and
$Y_L$ profiles shortly after bounce \citep[see also][]{Lentz_etal12}.
The spike of $Y_L$ is caused by electron neutrinos leaking out of the
shock.  The absorption of some of those neutrinos by the pre-shock
material results in a spike of $Y_e$.  In the post-shock region, the 
minimal $Y_e$ drops rapidly for the first few milliseconds after
bounce due to electron capture; it then becomes steady at $\sim 0.065
- 0.075$.  The velocity in some region behind the shock is positive
for the first $\sim$ 2 ms after bounce; the maximum positive velocity
reaches $\sim 2 \times 10^9\:\mathrm{cm}\;\mathrm{s}^{-1}$ similar to
the maximum positive velocity in \citet{ThompsonBP03}.  The maximum
temperature in the shocked region increases rapidly to
$17.5\:\mathrm{MeV}$ at $\sim 0.1\:\mathrm{ms}$ after bounce; it then
decreases quickly as the shock expands outward in radius; it increases
again and reaches $\sim 22\:\mathrm{MeV}$ at 200 ms after bounce due
to compression and deleptonization as matter settles onto the
proto-neutron star.  The evolution of the temperature, electron
fraction and velocity profiles is in agreement with that in previous
studies \citep[e.g.,][]{ThompsonBP03}.

\begin{figure}
\epsscale{0.5}
\plotone{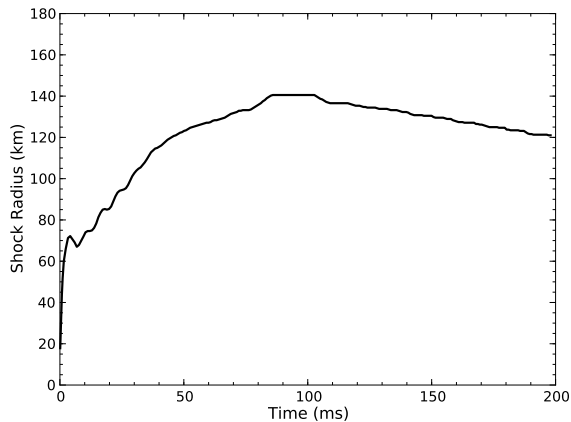}
\caption{ Shock radius vs. post-bounce time for the 1D core-collapse
  supernova simulation. The shock position is 
  defined as the place with the largest negative gradient of radial
  velocity.  The peak shock radius reaches $\sim 140\:\mathrm{km}$ at
  $\sim 100\:\mathrm{ms}$ after bounce.  
  \label{fig:rshock1d}}
\end{figure}

In Figure~\ref{fig:rshock1d}, we show shock radius as a function of
time.  For the first few milliseconds after bounce, the shock expands
very rapidly in radius, and reaches $\sim 72\:\mathrm{km}$.  The peak
shock radius reaches $\sim 140\:\mathrm{km}$ at $\sim
100\:\mathrm{ms}$ after bounce.  After that the shock slowly shrinks
in radius although it continues to grow in mass.

\begin{figure}
\epsscale{0.5}
\plotone{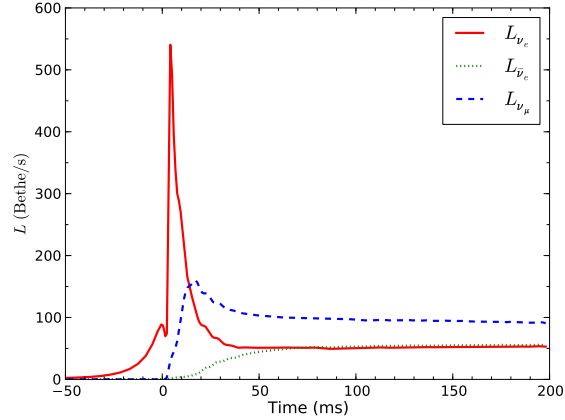}
\caption{ Comoving frame luminosities of $\nu_e$ ({\it red solid line}),
  $\bar{\nu}_e$ ({\it green dotted line}), and $\nu_\mu$ neutrinos
  ({\it blue dashed line}) measured at 500 km for the 1D core-collapse
  supernova simulation.  The luminosity of electron neutrinos reaches
  a peak value of $540\:\mathrm{Bethe}\:\mathrm{s}^{-1}$ at 4 ms after
  bounce.  For $L_{\nu_e}$, the full width at half maximum is $\sim
  7\:\mathrm{ms}$.  There is also a noticeable spike in the luminosity
  of $\nu_\mu$ neutrinos with a peak value of
  $160\:\mathrm{Bethe}\:\mathrm{s}^{-1}$ at $17\:\mathrm{ms}$ after
  bounce.
  \label{fig:L1d}}
\end{figure}

\begin{figure}
\epsscale{0.5}
\plotone{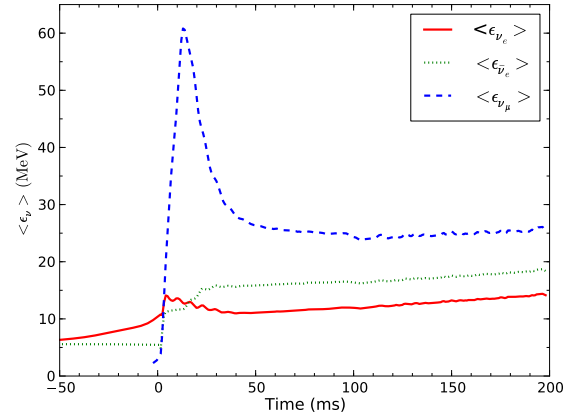}
\caption{ Rms average energies of $\nu_e$ ({\it red solid line}),
  $\bar{\nu}_e$ ({\it green dotted line}), and $\nu_\mu$ neutrinos
  ({\it blue dashed line}) measured at 500 km for the 1D core-collapse
  supernova simulation.  For $\langle\epsilon_{\nu_\mu}\rangle$, the
  peak value of $61\:\mathrm{MeV}$ is reached at 13 ms after
  bounce. The rms average energies at 200 ms after bounce are $14$,
  $19$, and $26\:\mathrm{MeV}$, for $\nu_e$, $\bar{\nu}_{e}$, and
  $\nu_\mu$, respectively.  We do not show $\langle\epsilon_{\nu_\mu}
  \rangle$ before bounce because there is very little $\nu_\mu$ at those
  early times.
  \label{fig:erms1d}}
\end{figure}

\begin{figure}
\epsscale{0.5}
\plotone{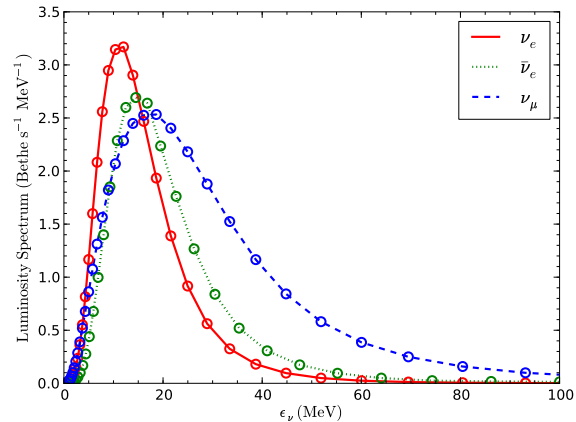}
\caption{ Luminosity spectrum of $\nu_e$ ({\it red solid line}),
  $\bar{\nu}_e$ ({\it green dotted line}), and $\nu_\mu$ neutrinos ({\it blue
    dashed line}) measured at 500 km and 200 ms after bounce for
  the 1D core-collapse supernova simulation.
  The actual energy groups are shown in circles. 
  \label{fig:spectra1d}}
\end{figure}

The comoving frame luminosities of $\nu_e$, $\bar{\nu}_e$, and
$\nu_\mu$ neutrinos measured at 500 km are shown in
Figure~\ref{fig:L1d}.  In agreement with previous studies
\citep[e.g.,][]{ThompsonBP03, MarekJanka09, Lentz_etal12}, there is a
small dip in $L_{\nu_e}$ after the initial rise.  The dip is followed
by a large pulse in $L_{\nu_e}$ that reaches
$540\:\mathrm{Bethe}\:\mathrm{s}^{-1}$ at 4 ms after bounce.  The full
width at half maximum for $L_{\nu_e}$ is $\sim 7\:\mathrm{ms}$.  The
peak $L_{\nu_e}$ in our model is about 20\% larger than that in
\citet{MarekJanka09, Lentz_etal12}.  In our results, there is also a
noticeable spike in $L_{\nu_\mu}$.  Note that \citet{ThompsonBP03,
  Lentz_etal12} have obtained somewhat similar spikes in their
simulations that did not include inelastic scattering.  The lack of
inelastic scattering of neutrinos on electrons results in harder
neutrino spectra, especially for $\nu_\mu$, as shown in the rms
average energies (Figure~\ref{fig:erms1d}).  Here the rms average
is defined as
\begin{equation}
  \langle \epsilon_\nu \rangle = \left[ \frac{\int \epsilon_\nu^2 n_\nu
      \mathrm{d} \nu}{\int n_\nu \mathrm{d} \nu} \right]^{1/2}, 
\end{equation}
where $\epsilon_\nu$ is the neutrino energy and $n_\nu$ is the
monochromatic neutrino number density.  Nevertheless we obtain
$\langle \epsilon_{\nu_e} \rangle <
\langle\epsilon_{\bar{\nu}_e}\rangle < \langle \epsilon_{\nu_\mu}
\rangle$ after the initial phases, as expected.  At 200 ms after
bounce, we obtain $\langle \epsilon_{\nu_e} \rangle \approx 14$,
$\langle\epsilon_{\bar{\nu}_e}\rangle \approx 19$, and $\langle
\epsilon_{\nu_\mu} \rangle \approx 26\:\mathrm{MeV}$.  The luminosity
spectra measured at 500 km in the comoving are shown
Figure~\ref{fig:spectra1d}.  The spectra are evidently harder than
those in \citet{ThompsonBP03} because inelastic scattering is not
included in our model. 

\subsubsection{Two-Dimensional Core-Collapse Supernova Simulation}

Here we present a two-dimensional core-collapse supernova simulation.
The computational domain for this 2D cylindrical run is $0<r<2.5\times
10^8\,\mathrm{cm}$ and $-2.5\times 10^8\,\mathrm{cm} <z<2.5\times
10^8\,\mathrm{cm}$.  Two refinement levels (three total levels) with a
refinement factor of 4 are used.  The coarsest level has 256 and 512
cells at $r$-direction and $z$-direction, respectively.  Thus the size
of the cells on the finest level is about $0.61\,\mathrm{km}$. The
same refinement criteria as those in the 1D simulation in
Section~\ref{sec:ccsn1d} are used.  We use 20, 16, and 20 groups with
logarithmic spacing for $\nu_e$, $\bar{\nu}_e$, and $\nu_\mu$,
respectively.  The other aspects of this 2D run such as the outer
boundary and flux limiter are the same as those in the 1D simulation
in Section~\ref{sec:ccsn1d}.  We start the 2D simulation from the
result of a 1D simulation rather than the $15\,M_\odot$ presupernova
model.  The initial model is obtained from the 1D simulation at about
10 ms before bounce.  At that time, multi-dimensional effects are
expected to still be small.  The 1D simulation that provides the
initial model for the 2D run has similar resolution as the 2D run.
Note that, other than resolution, the low-resolution 1D simulation here
is essentially the same as the 1D high-resolution simulation in
Section~\ref{sec:ccsn1d}.

\begin{figure}
\epsscale{0.7}
\plotone{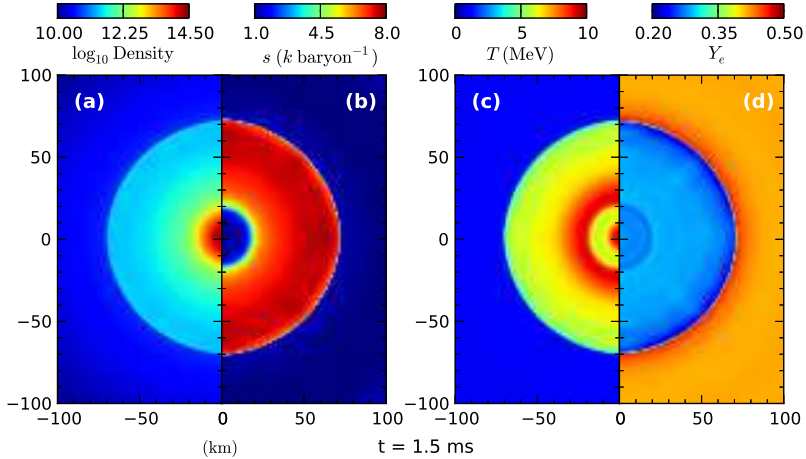}
\caption{ Snapshot at 1.5 ms after bounce for the 2D cylindrical core-collapse
  supernova simulation.  We show (a) logarithmic density, (b) entropy
  per baryon, (c) temperature, and (d) electron fraction, where the
  unit for density is $\mathrm{g}\:\mathrm{cm}^{-3}$.  Only the inner
  100 km of the model is shown here.   
  \label{fig:c2d1}}
\end{figure}

\begin{figure}
\epsscale{0.7}
\plotone{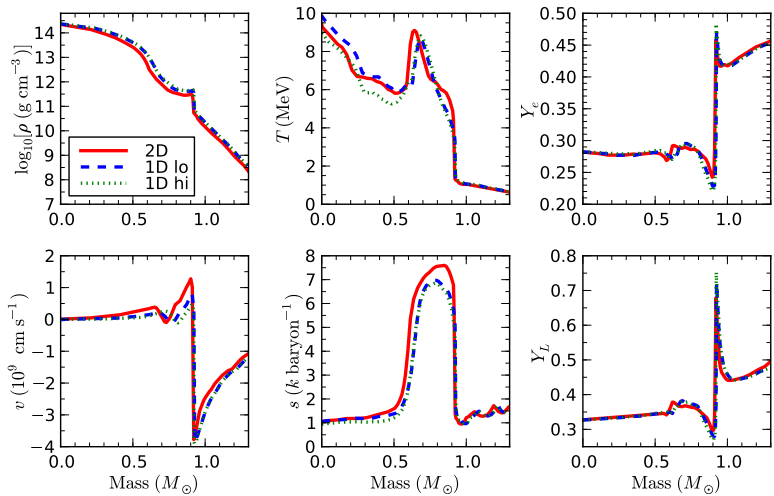}
\caption{ Density ({\it upper left}), radial velocity ({\it lower left}),
  temperature ({\it upper center}), entropy per baryon ({\it lower
    center}), electron fraction ({\it upper right}), and net lepton
  fraction ({\it lower right}) as a function of enclosed mass 
  at 1.5 ms after bounce.    Here, radial velocity is the velocity in
  spherical radial direction.  We show the results of the 2D
  cylindrical simulation ({\it red solid lines}), 1D low-resolution
  spherical run ({\it blue dashed lines}), and 1D high-resolution
  spherical run ({\it green dotted lines}).  The radial profiles of the
  2D results are generated via averaging. 
  \label{fig:ss2d1}}
\end{figure}

\begin{figure}
\epsscale{0.67}
\plotone{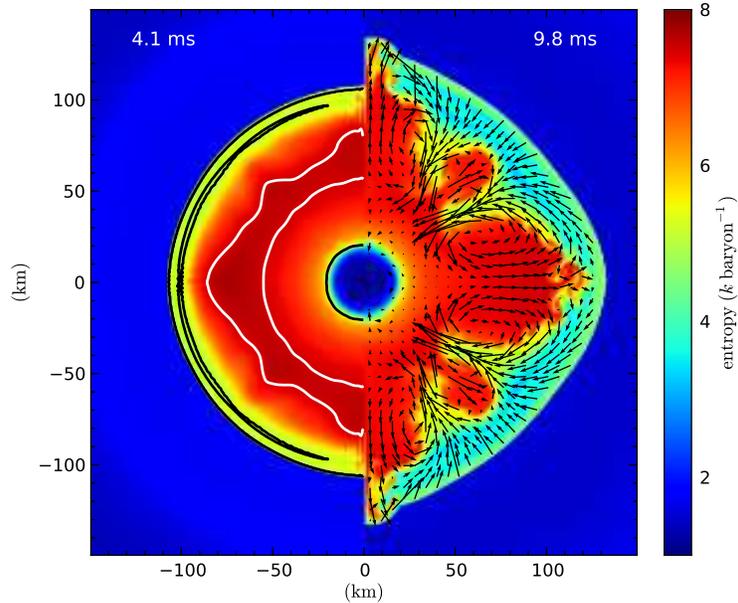}
\caption{ Entropy at 4.1 ({\it left}) and 9.8 ms ({\it right}) after
  bounce for the 2D cylindrical core-collapse 
  supernova simulation.  The white and black contour lines in the left panel
  denote $s = 7.5$ and $5\:k\:\mathrm{baryon}^{-1}$, respectively.
  The arrows in the right panel show the velocity field in the shocked
  region at 9.8 ms after bounce.  It is shown 
  that convection has developed in the shocked region at 9.8 ms
  after bounce.
  \label{fig:conv1}}
\end{figure}

\begin{figure}
\epsscale{0.7}
\plotone{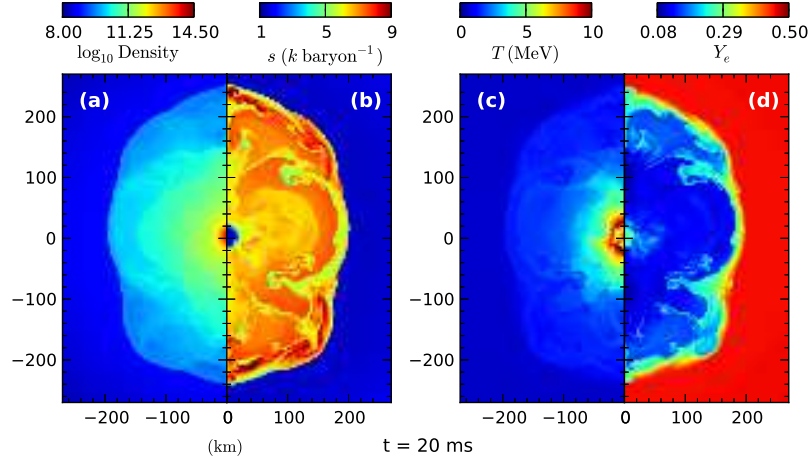}
\caption{ Snapshot at 20 ms after bounce for the 2D cylindrical core-collapse
  supernova simulation.  We show (a) logarithmic density, (b) entropy
  per baryon, (c) temperature, and (d) electron fraction, where the
  unit for density is $\mathrm{g}\:\mathrm{cm}^{-3}$.  Only the inner
  270 km of the model is shown here. 
  \label{fig:c2d2}}
\end{figure}

\begin{figure}
\epsscale{0.5}
\plotone{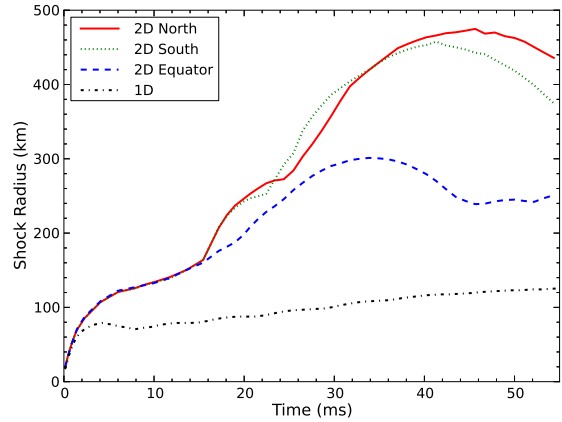}
\caption{ Shock radius vs. post-bounce time for the 2D core-collapse
  supernova simulation.  We show the shock position in the northern
  hemisphere ({\it red solid 
    line}), southern hemisphere ({\it green dotted line}), and equatorial
  plane ({\it blue dashed lines}) for the 2D simulation.  Also shown
  is the result of the 1D low-resolution simulation ({\it black
    dash-dotted line}).  
  \label{fig:rshock2d}}
\end{figure}

\begin{figure}
\epsscale{0.5}
\plotone{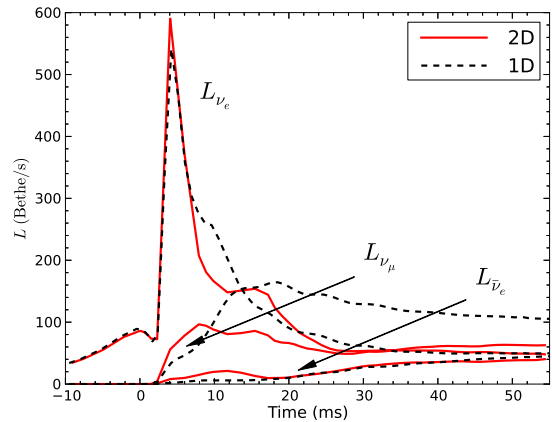}
\caption{ Comoving frame luminosities of $\nu_e$, $\bar{\nu}_e$, and
  $\nu_\mu$ neutrinos measured at 500 km for the 2D cylindrical ({\it
    red solid lines}) and the low resolution 1D spherical ({\it black
    dashed lines}) core-collapse supernova simulations.  
  \label{fig:L2d}}
\end{figure}

Figure~\ref{fig:c2d1} shows the density, entropy, temperature, and
electron fraction profiles at 1.5 ms after bounce.  The profiles are
still highly symmetric.  Spherically averaged profiles of density,
velocity, temperature, entropy, electron fraction, and lepton fraction
at 1.5 ms are shown in Figure~\ref{fig:ss2d1}.  Also shown in
Figure~\ref{fig:ss2d1} are the results of the two 1D runs.  At this
point, the 2D results are only slightly different from the 1D results.
However, in some region behind the shock, the gradient of entropy is
negative.  A negative entropy gradient region from $\sim
70\,\mathrm{km}$ to $\sim 100\,\mathrm{km}$ is also clearly visible in
the entropy profile at 4.1 ms (left panel of Figure~\ref{fig:conv1}).
The white and black contour lines in the left panel of
Figure~\ref{fig:conv1} denote $s = 7.5$ and
$5\:k\:\mathrm{baryon}^{-1}$, respectively.  Convective instabilities
can potentially develop in the region between $\sim 60\,\mathrm{km}$
and $90\,\mathrm{km}$.  In fact, convection does emerge quickly in our
simulation.  In the right panel of Figure~\ref{fig:conv1}, we show the
entropy profile at 9.8 ms after bounce.  Also shown in the figure is
the velocity field.  High entropy plumes rise while some low entropy
material falls down.  Secondary Kelvin-Helmholtz instabilities also
develop in the shearing regions.  The convection helps push the shock
outwards because of the enhanced material energy transport from the
inner hot region to the region behind the shock by the convective
motion.  Figure~\ref{fig:c2d2} shows the density, entropy,
temperature, and electron fraction profiles at 20 ms after bounce.
The shock has moved to $\sim 200\,\mathrm{km}$ at 20 ms in the 2D
simulation (see also Figure~\ref{fig:rshock2d}).  At 40 ms after
bounce, the north pole and the south pole of the shock have moved to
$\sim 450\,\mathrm{km}$.  In contrast, in the 1D simulation in
Section~\ref{sec:ccsn1d} (Figure~\ref{fig:rshock1d}), the shock radius
is $\sim 85\,\mathrm{km}$ at 20 ms and it never exceeds
$141\,\mathrm{km}$.  The comoving frame luminosities of $\nu_e$,
$\bar{\nu}_e$, and $\nu_\mu$ neutrinos measured at 500 km for the 2D
simulation are shown in Figure~\ref{fig:L2d}.  Also shown in
Figure~\ref{fig:L2d} are the results of the 1D low-resolution run,
which are nearly identical to those of the 1D high-resolution results
(Figure~\ref{fig:L1d}).  In the 2D results, there is also a large
narrow pulse in $L_{\nu_e}$ that rises and drops at roughly the same
rate.  However, after $\sim 7\:\mathrm{ms}$ past bounce, the 2D
results are qualitatively different from the 1D results because of the
emergence of the convection.  A major difference is that $\nu_\mu$
neutrinos are much less luminous in 2D than 1D, because the
temperature in 2D is cooler than that in 1D due to the rapid expansion
of 2D shock.

Since this paper is not on core-collapse supernova explosions per se,
the 2D simulation is stopped at $\sim 55\:\mathrm{ms}$ after bounce.
A detailed analysis of the 2D simulation is beyond the scope of this
paper on algorithm, implementation, and testing.

\section{Parallel Performance}
\label{sec:performance}

\begin{figure}
\epsscale{0.5}
\plotone{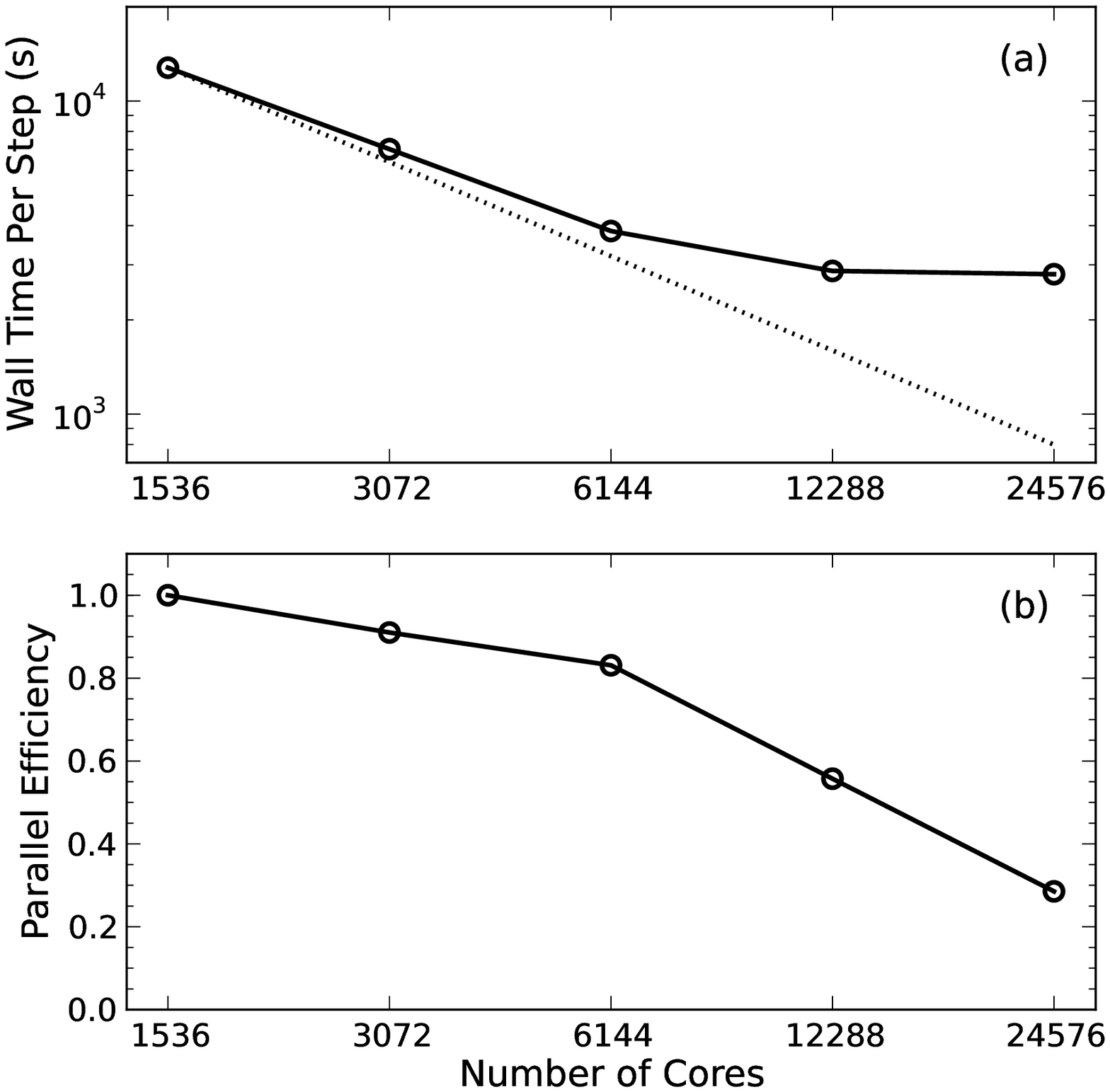}
\caption{ Strong scaling behavior of 3D simulations on Hopper at NERSC. 
  We show (a) wall clock time per coarse time step and (b) parallel
  efficiency in circles for five runs with the number of cores ranging from 1536
  to 24576.  The dotted line in panel (a) indicates perfect scaling.  
  \label{fig:scaling}}
\end{figure}

In paper II, we demonstrated that the gray radiation solver in
CASTRO has very good scaling behavior on up to 32768 cores via a weak
scaling study.  The multigroup radiation solver presented in this
paper has similar good weak scaling behavior because the same module
for solving Equation~(\ref{eq:canon}) is used in both solvers.  Here
we present a strong scaling study that was carried out on Hopper, a
petascale Cray XE6 supercomputer at the National Energy Research
Scientific Computing Center.  We have performed a series of
three-dimensional core-collapse supernova simulations in Cartesian
coordinates with the same resolution on various numbers of cores.  A
hybrid MPI/OpenMP approach with 6 OpenMP threads per MPI task was used
for these simulations.  PFMG, a parallel multigrid solver from {\it
  hypre} \citep{hypre,hypreweb}, was used to solve the linear systems.
In each run, we start the 3D simulation from the result of a 1D
simulation at about 10 ms before bounce.  We use 12 energy groups for
each species ($\nu_e$, $\bar{\nu}_e$, and $\nu_\mu$).  The
computational domain is a three-dimensional box with a size of
$5000\:\mathrm{km}$ in each dimension.  Three refinement levels (four
total levels) are used with a refinement factor of 2 between level 0
and level 1, and a refinement factor of 4 for other levels.  The
coarsest level has $256^3$ cells.  The finest level, which
approximately covers the inner region of $\sqrt{x^2+y^2+z^2} <
165\;\mathrm{km}$, has a cell size of $0.61\:\mathrm{km}$.
Figure~\ref{fig:scaling} shows the wall clock time per coarse time
step for a series of runs with the number of cores ranging from 1536
to 24576. Also shown in Figure~\ref{fig:scaling} is parallel efficient
defined as
\begin{equation}
  E(N) = \frac{N_0 T_0}{N T_N},
\end{equation}
where $N$ is the number of cores, $T_N$ is the wall clock time for the
$N$-core run, $N_0$ is the number of cores in a baseline run (i.e.,
1536-core run), and $T_0$ is the wall clock time for the baseline run.
In this strong scaling study, good scaling behavior is achieved on up
to $\sim 10000$ cores.  However, because this is a strong scaling
study, the work load per core diminishes as more
cores are used.  For example, the average number of cells for each
core in the 24576-core run is about $16^3$.  An unfavorable
consequence of small computational boxes is an increased communication
overhead.  Another unfavorable consequence is that more computation is
wasted at domain boundaries due to larger surface area to volume
ratios of smaller boxes.  Hence, the degradation of parallel
performance on more than 10000 cores is not surprising.

\section{Summary}
\label{sec:sum}

In this paper, we have presented a multi-dimensional multigroup
radiation hydrodynamics solver that is part of the CASTRO code.
Block-structured AMR is utilized in CASTRO.  In this paper, we focus
attention on the single-level algorithms because the AMR algorithms
have been presented in Papers I, II, and references therein.

Our multigroup radiation hydrodynamics solver is based on a comoving
frame formulation that is correct to order $O(v/c)$, and uses a FLD
approximation.  Our mathematical analysis shows that the system we
solve contains a hyperbolic subsystem that is the usual hydrodynamics
system modified by radiation, a frequency space advection part that
accounts for the Doppler shift of radiation due to the motion of
matter, and a parabolic diffusion part.  We have presented the
mathematical characteristics of the hyperbolic subsystem.  The
eigenvalues and eigenvectors we have obtained are useful for
characteristic-based Godunov schemes.  We also described our treatment
of the frequency space advection part.  An implicit solver involving
two nested iterations is presented.  We have also presented two
acceleration schemes that improve the convergence rate of the implicit
solver.

Extensive testing is presented demonstrating the accuracy and
stability of CASTRO to solve multigroup radiation hydrodynamics
problems.  We have presented a series of photon radiation test
problems that cover a wide range of regimes.  In addition to photon
radiation problems, we have demonstrated the applicability of CASTRO
in core-collapse supernova simulations.

Recently, \citet{Lentz_etal12} have argued, via a series of 1D
simulations, that general relativistic effects and inelastic
scattering of neutrinos on matter have significant effects on the
numerical modeling of core-collapse supernova explosions.  In
contrast, \citet{NordhausBAB10} have concluded that spatial
dimensionality has far more impact than general relativistic effects
or microphysics such as inelastic scattering.  Our 1D simulations are
consistent with the assessment of \citet{Lentz_etal12} on the effects
of inelastic scattering.  However, our 2D simulation supports the
assessment of \citet{NordhausBAB10} on the importance of spatial
dimensionality.  Our simulations have shown that multidimensional
effects (e.g., convective motion in the region behind the shock and
exterior to the nascent neutron star) appear less than 10 milliseconds
after bounce, and 2D results are profoundly different from 1D results.
Thus, one should be cautious about assessment of various effects
such as inelastic scattering that are based on 1D simulations that last
several hundred milliseconds past bounce like the ones presented in
\citet{Lentz_etal12}.

The main advantages of the CASTRO code are the efficiency due to the
use of AMR and the accuracy due to the coupling of radiation force
into the Riemann solver.  Three-dimensional multigroup
neutrino radiation hydrodynamics simulations of core-collapse
supernovae with a good resolution using CASTRO can be carried out,
if not now, in the near future with faster computers.  In the strong
scaling study with a total of 36 radiation groups and a cell size of
$0.6\:\mathrm{km}$ on the finest level
(Section~\ref{sec:performance}), it took about 0.8 hours to advance
one coarse time step (about $0.1\:\mathrm{ms}$) on 12288 cores of
Hopper.  Thus, it would take 800 hours on 12288 cores of Hopper to
evolve to $100\:\mathrm{ms}$.  Admittedly, it is still very expensive
to perform long-duration 3D simulations, but it is possible to study
the first $100\:\mathrm{ms}$ after bounce and investigate the
development of turbulent convection region via 3D simulations.

\acknowledgments

The work at LBNL was supported by the Office of High Energy Physics
and the Office of Advanced Scientific Computing Research of the
U.S. Department of Energy under contract No.\ DE-AC02-05CH11231.  The
work performed at LLNL was supported by the SciDAC program of the
U.S. Department of Energy under the auspices of contract No.\
DE-AC52-07NA27344.  Adam Burrows was supported by the SciDAC program
of DOE under grant number DE-FG02-08ER41544, the NSF under subaward
no. ND201387 to the Joint Institute for Nuclear Astrophysics, and the
NSF PetaApps program, under award OCI-0905046 via a subaward no. 44592
from Louisiana State University to Princeton University.  This
research used resources of the National Energy Research Scientific
Computing Center, which is supported by the Office of Science of the
U.S.  Department of Energy under Contract No. DE-AC02-05CH11231.  The
authors would like to thank S.E. Woosley for many useful comments on
the manuscript.


\begin{thebibliography}{}

\bibitem[{{Abdikamalov} {et~al.}(2012){Abdikamalov}, {Burrows}, {Ott},
  {L{\"o}ffler}, {O'Connor}, {Dolence}, \& {Schnetter}}]{AbdikamalovBO12}
{Abdikamalov}, E., {Burrows}, A., {Ott}, C.~D., {L{\"o}ffler}, F., {O'Connor},
  E., {Dolence}, J.~C., \& {Schnetter}, E. 2012, \apj, accepted

\bibitem[{{Alme} \& {Wilson}(1973)}]{AlmeWilson73}
{Alme}, M.~L., \& {Wilson}, J.~R. 1973, \apj, 186, 1015

\bibitem[{{Almgren} {et~al.}(1998){Almgren}, {Bell}, {Colella}, {Howell}, \&
  {Welcome}}]{IAMR}
{Almgren}, A., {Bell}, J.~B., {Colella}, P., {Howell}, L.~H., \& {Welcome},
  M.~L. 1998, Journal of Computational Physics, 142, 1

\bibitem[{{Almgren} {et~al.}(2010){Almgren}, {Beckner}, {Bell}, {Day},
  {Howell}, {Joggerst}, {Lijewski}, {Nonaka}, {Singer}, \& {Zingale}}]{CASTRO}
{Almgren}, A.~S. {et~al.} 2010, \apj, 715, 1221

\bibitem[{{Bell} {et~al.}(1989){Bell}, {Colella}, \&
  {Trangenstein}}]{bellcolellatrangenstein}
{Bell}, J.~B., {Colella}, P., \& {Trangenstein}, J.~A. 1989, Journal of
  Computational Physics, 82, 362

\bibitem[{{Bethe}(1990)}]{Bethe90}
{Bethe}, H.~A. 1990, Reviews of Modern Physics, 62, 801

\bibitem[{{Bowers} \& {Wilson}(1982)}]{BowersWilson82}
{Bowers}, R.~L., \& {Wilson}, J.~R. 1982, \apjs, 50, 115

\bibitem[{{Bruenn}(1985)}]{Bruenn85}
{Bruenn}, S.~W. 1985, \apjs, 58, 771

\bibitem[{{Bruenn} {et~al.}(1978){Bruenn}, {Buchler}, \& {Yueh}}]{BruennBY78}
{Bruenn}, S.~W., {Buchler}, J.~R., \& {Yueh}, W.~R. 1978, \apss, 59, 261

\bibitem[{{Buchler}(1983)}]{Buchler83}
{Buchler}, J.~R. 1983, \jqsrt, 30, 395

\bibitem[{{Buras} {et~al.}(2006){Buras}, {Rampp}, {Janka}, \&
  {Kifonidis}}]{BurasRJK06}
{Buras}, R., {Rampp}, M., {Janka}, H.-T., \& {Kifonidis}, K. 2006, \aap, 447,
  1049

\bibitem[{{Burrows} {et~al.}(1995){Burrows}, {Hayes}, \&
  {Fryxell}}]{BurrowsHF95}
{Burrows}, A., {Hayes}, J., \& {Fryxell}, B.~A. 1995, \apj, 450, 830

\bibitem[{{Burrows} {et~al.}(2007){Burrows}, {Livne}, {Dessart}, {Ott}, \&
  {Murphy}}]{BurrowsLD07}
{Burrows}, A., {Livne}, E., {Dessart}, L., {Ott}, C.~D., \& {Murphy}, J. 2007,
  \apj, 655, 416

\bibitem[{{Burrows} {et~al.}(2006){Burrows}, {Reddy}, \&
  {Thompson}}]{BurrowsRT06}
{Burrows}, A., {Reddy}, S., \& {Thompson}, T.~A. 2006, Nuclear Physics A, 777,
  356

\bibitem[{{Burrows} {et~al.}(2000){Burrows}, {Young}, {Pinto}, {Eastman}, \&
  {Thompson}}]{BurrowsYP00}
{Burrows}, A., {Young}, T., {Pinto}, P., {Eastman}, R., \& {Thompson}, T.~A.
  2000, \apj, 539, 865

\bibitem[{{Castor}(2004)}]{Castor04}
{Castor}, J.~I. 2004, {Radiation Hydrodynamics} (Cambridge, UK: Cambridge
  University Press)

\bibitem[{{Clark}(1987)}]{Clark87}
{Clark}, B.~A. 1987, Journal of Computational Physics, 70, 311

\bibitem[{{Colella} {et~al.}(1997){Colella}, {Glaz}, \&
  {Ferguson}}]{ColellaGF97}
{Colella}, P., {Glaz}, H.~M., \& {Ferguson}, R.~E. 1997, unpublished manuscript

\bibitem[{Eisenstat(1981)}]{HLLE}
Eisenstat, S.~C. 1981, SIAM Journal on Scientific and Statistical Computing, 2,
  1

\bibitem[{Falgout \& Yang(2002)}]{hypre}
Falgout, R., \& Yang, U. 2002, in Lecture Notes in Computer Science, Vol. 2331,
  Computational Science — ICCS 2002, ed. P.~Sloot, A.~Hoekstra, C.~Tan, \&
  J.~Dongarra (Springer Berlin / Heidelberg), 632--641

\bibitem[{{Fryer} {et~al.}(2006){Fryer}, {Rockefeller}, \&
  {Warren}}]{FryerRW06}
{Fryer}, C.~L., {Rockefeller}, G., \& {Warren}, M.~S. 2006, \apj, 643, 292

\bibitem[{{Fryer} \& {Warren}(2002)}]{FryerWarren02}
{Fryer}, C.~L., \& {Warren}, M.~S. 2002, \apjl, 574, L65

\bibitem[{{Gittings} {et~al.}(2008){Gittings}, {Weaver}, {Clover}, {Betlach},
  {Byrne}, {Coker}, {Dendy}, {Hueckstaedt}, {New}, {Oakes}, {Ranta}, \&
  {Stefan}}]{RAGE}
{Gittings}, M. {et~al.} 2008, Computational Science and Discovery, 1, 015005

\bibitem[{Graziani(2008)}]{Graziani08}
Graziani, F. 2008, in Lecture Notes in Computational Science and Engineering,
  Vol.~62, Computational Methods in Transport: Verification and Validation, ed.
  F.~Graziani (Springer Berlin Heidelberg), 151--167

\bibitem[{{Harten} {et~al.}(1983){Harten}, {Lax}, \& {van Leer}}]{HLL}
{Harten}, A., {Lax}, P.~D., \& {van Leer}, B. 1983, SIAM Review, 25, 35

\bibitem[{{Herant} {et~al.}(1994){Herant}, {Benz}, {Hix}, {Fryer}, \&
  {Colgate}}]{HerantBH94}
{Herant}, M., {Benz}, W., {Hix}, W.~R., {Fryer}, C.~L., \& {Colgate}, S.~A.
  1994, \apj, 435, 339

\bibitem[{{Hubeny} \& {Burrows}(2007)}]{HubenyBurrows07}
{Hubeny}, I., \& {Burrows}, A. 2007, \apj, 659, 1458

\bibitem[{{{\it hypre} Code Project}(2012)}]{hypreweb}
{{\it hypre} Code Project}. 2012, \texttt{http://www.llnl.gov/CASC/hypre/}

\bibitem[{Janka(2012)}]{Janka12}
Janka, H.-T. 2012, Annual Review of Nuclear and Particle Science, 62, in press

\bibitem[{{Janka} {et~al.}(2007){Janka}, {Langanke}, {Marek},
  {Mart{\'{\i}}nez-Pinedo}, \& {M{\"u}ller}}]{JankaLM07}
{Janka}, H.-T., {Langanke}, K., {Marek}, A., {Mart{\'{\i}}nez-Pinedo}, G., \&
  {M{\"u}ller}, B. 2007, \physrep, 442, 38

\bibitem[{{Kershaw}(1976)}]{Kershaw76}
{Kershaw}, D.~S. 1976, {Flux Limiting Nature's Own Way}, Tech. Rep. UCRL-78378,
  Lawrence Livermore National Laboratory

\bibitem[{{Kotake} {et~al.}(2006){Kotake}, {Sato}, \& {Takahashi}}]{KotakeST06}
{Kotake}, K., {Sato}, K., \& {Takahashi}, K. 2006, Reports on Progress in
  Physics, 69, 971

\bibitem[{{Krumholz} {et~al.}(2007){Krumholz}, {Klein}, {McKee}, \&
  {Bolstad}}]{KrumholzKMB07}
{Krumholz}, M.~R., {Klein}, R.~I., {McKee}, C.~F., \& {Bolstad}, J. 2007, APJ,
  667, 626

\bibitem[{{Lentz} {et~al.}(2012){Lentz}, {Mezzacappa}, {Bronson Messer},
  {Liebend{\"o}rfer}, {Hix}, \& {Bruenn}}]{Lentz_etal12}
{Lentz}, E.~J., {Mezzacappa}, A., {Bronson Messer}, O.~E., {Liebend{\"o}rfer},
  M., {Hix}, W.~R., \& {Bruenn}, S.~W. 2012, \apj, 747, 73

\bibitem[{LeVeque(2002)}]{LeVeque}
LeVeque, R.~J. 2002, Finite-Volume Methods for Hyperbolic Problems (Cambridge
  University Press)

\bibitem[{{Levermore}(1984)}]{Levermore84}
{Levermore}, C.~D. 1984, Journal of Quantitative Spectroscopy and Radiative
  Transfer, 31, 149

\bibitem[{{Levermore} \& {Pomraning}(1981)}]{LevermorePomraning81}
{Levermore}, C.~D., \& {Pomraning}, G.~C. 1981, \apj, 248, 321

\bibitem[{{Liebend{\"o}rfer} {et~al.}(2004){Liebend{\"o}rfer}, {Messer},
  {Mezzacappa}, {Bruenn}, {Cardall}, \& {Thielemann}}]{LiebendorferMM04}
{Liebend{\"o}rfer}, M., {Messer}, O.~E.~B., {Mezzacappa}, A., {Bruenn}, S.~W.,
  {Cardall}, C.~Y., \& {Thielemann}, F.-K. 2004, \apjs, 150, 263

\bibitem[{{Liebend{\"o}rfer} {et~al.}(2001){Liebend{\"o}rfer}, {Mezzacappa},
  {Thielemann}, {Messer}, {Hix}, \& {Bruenn}}]{Liebendorfer_etal01}
{Liebend{\"o}rfer}, M., {Mezzacappa}, A., {Thielemann}, F.-K., {Messer}, O.~E.,
  {Hix}, W.~R., \& {Bruenn}, S.~W. 2001, \prd, 63, 103004

\bibitem[{{Livne} {et~al.}(2004){Livne}, {Burrows}, {Walder}, {Lichtenstadt},
  \& {Thompson}}]{LivneBW04}
{Livne}, E., {Burrows}, A., {Walder}, R., {Lichtenstadt}, I., \& {Thompson},
  T.~A. 2004, \apj, 609, 277

\bibitem[{{Lowrie} \& {Edwards}(2008)}]{LowrieEdwards08}
{Lowrie}, R.~B., \& {Edwards}, J.~D. 2008, Shock Waves, 18, 129

\bibitem[{{Marek} \& {Janka}(2009)}]{MarekJanka09}
{Marek}, A., \& {Janka}, H.-T. 2009, \apj, 694, 664

\bibitem[{{Marinak} {et~al.}(2001){Marinak}, {Kerbel}, {Gentile}, {Jones},
  {Munro}, {Pollaine}, {Dittrich}, \& {Haan}}]{HYDRA}
{Marinak}, M.~M., {Kerbel}, G.~D., {Gentile}, N.~A., {Jones}, O., {Munro}, D.,
  {Pollaine}, S., {Dittrich}, T.~R., \& {Haan}, S.~W. 2001, Physics of Plasmas,
  8, 2275

\bibitem[{{Mezzacappa} \& {Bruenn}(1993)}]{MezzacappaBruenn93}
{Mezzacappa}, A., \& {Bruenn}, S.~W. 1993, \apj, 405, 669

\bibitem[{{Mihalas} \& {Mihalas}(1999)}]{MihalasMihalas99}
{Mihalas}, D., \& {Mihalas}, B.~W. 1999, {Foundations of radiation
  hydrodynamics} (New York: Dover)

\bibitem[{{Miller} {et~al.}(1993){Miller}, {Wilson}, \& {Mayle}}]{Millerwm93}
{Miller}, D.~S., {Wilson}, J.~R., \& {Mayle}, R.~W. 1993, \apj, 415, 278

\bibitem[{{Miller} \& {Colella}(2002)}]{unsplitPPM}
{Miller}, G.~H., \& {Colella}, P. 2002, Journal of Computational Physics, 183,
  26

\bibitem[{{Minerbo}(1978)}]{Minerbo78}
{Minerbo}, G.~N. 1978, \jqsrt, 20, 541

\bibitem[{{Morel} {et~al.}(1985){Morel}, {Larsen}, \& {Matzen}}]{MorelLM85}
{Morel}, J., {Larsen}, E.~W., \& {Matzen}, M.~K. 1985, \jqsrt, 34, 243

\bibitem[{{Morel} {et~al.}(2007){Morel}, {Brian Yang}, \& {Warsa}}]{MorelYW07}
{Morel}, J.~E., {Brian Yang}, T.-Y., \& {Warsa}, J.~S. 2007, Journal of
  Computational Physics, 227, 244

\bibitem[{{Nordhaus} {et~al.}(2010){Nordhaus}, {Burrows}, {Almgren}, \&
  {Bell}}]{NordhausBAB10}
{Nordhaus}, J., {Burrows}, A., {Almgren}, A., \& {Bell}, J. 2010, \apj, 720,
  694

\bibitem[{{O'Connor} \& {Ott}(2010)}]{OConnorOtt10}
{O'Connor}, E., \& {Ott}, C.~D. 2010, Classical and Quantum Gravity, 27, 114103

\bibitem[{{Ott} {et~al.}(2008){Ott}, {Burrows}, {Dessart}, \&
  {Livne}}]{OttBDL08}
{Ott}, C.~D., {Burrows}, A., {Dessart}, L., \& {Livne}, E. 2008, \apj, 685,
  1069

\bibitem[{{Rampp} \& {Janka}(2000)}]{RamppJanka00}
{Rampp}, M., \& {Janka}, H.-T. 2000, \apjl, 539, L33

\bibitem[{{Rampp} \& {Janka}(2002)}]{RamppJanka02}
---. 2002, \aap, 396, 361

\bibitem[{{Sanchez} \& {Pomraning}(1991)}]{SanchezPomraning91}
{Sanchez}, R., \& {Pomraning}, G.~C. 1991, \jqsrt, 45, 313

\bibitem[{{Shen} {et~al.}(1998{\natexlab{a}}){Shen}, {Toki}, {Oyamatsu}, \&
  {Sumiyoshi}}]{SHenEOSa}
{Shen}, H., {Toki}, H., {Oyamatsu}, K., \& {Sumiyoshi}, K. 1998{\natexlab{a}},
  Nuclear Physics A, 637, 435

\bibitem[{{Shen} {et~al.}(1998{\natexlab{b}}){Shen}, {Toki}, {Oyamatsu}, \&
  {Sumiyoshi}}]{ShenEOSb}
---. 1998{\natexlab{b}}, Progress of Theoretical Physics, 100, 1013

\bibitem[{{Shestakov} \& {Bolstad}(2005)}]{ShestakovBolstab05}
{Shestakov}, A.~I., \& {Bolstad}, J.~H. 2005, \jqsrt, 91, 133

\bibitem[{Shestakov \& Offner(2008)}]{ShestakovOffner08}
Shestakov, A.~I., \& Offner, S. S.~R. 2008, Journal of Computational Physics,
  227, 2154

\bibitem[{{Shu} \& {Osher}(1988)}]{ShuOsher88}
{Shu}, C.-W., \& {Osher}, S. 1988, Journal of Computational Physics, 77, 439

\bibitem[{{Su} \& {Olson}(1999)}]{SuOlson99}
{Su}, B., \& {Olson}, G.~L. 1999, \jqsrt, 62, 279

\bibitem[{{Swesty} \& {Myra}(2009)}]{SwestyMyra09}
{Swesty}, F.~D., \& {Myra}, E.~S. 2009, \apjs, 181, 1

\bibitem[{{Takiwaki} {et~al.}(2012){Takiwaki}, {Kotake}, \&
  {Suwa}}]{TakiwakiKS12}
{Takiwaki}, T., {Kotake}, K., \& {Suwa}, Y. 2012, \apj, 749, 98

\bibitem[{{Thompson} {et~al.}(2003){Thompson}, {Burrows}, \&
  {Pinto}}]{ThompsonBP03}
{Thompson}, T.~A., {Burrows}, A., \& {Pinto}, P.~A. 2003, \apj, 592, 434

\bibitem[{{van der Holst} {et~al.}(2011){van der Holst}, {T{\'o}th}, {Sokolov},
  {Powell}, {Holloway}, {Myra}, {Stout}, {Adams}, {Morel}, {Karni}, {Fryxell},
  \& {Drake}}]{crash}
{van der Holst}, B. {et~al.} 2011, \apjs, 194, 23

\bibitem[{{van Leer}(1977)}]{MC}
{van Leer}, B. 1977, Journal of Computational Physics, 23, 276

\bibitem[{{Woosley} \& {Weaver}(1995)}]{WoosleyWeaver95}
{Woosley}, S.~E., \& {Weaver}, T.~A. 1995, \apjs, 101, 181

\bibitem[{{Zhang} {et~al.}(2011){Zhang}, {Howell}, {Almgren}, {Burrows}, \&
  {Bell}}]{CASTRO2}
{Zhang}, W., {Howell}, L., {Almgren}, A., {Burrows}, A., \& {Bell}, J. 2011,
  \apjs, 196, 20

\end{thebibliography}
\end{document}